\pdfoutput=1

\documentclass[aps,pra,reprint,superscriptaddress,longbibliography]{revtex4-2}

\usepackage[T1]{fontenc}
\usepackage{lmodern}
\usepackage{bm}
\usepackage{microtype}
\usepackage{amsmath,amssymb,amsfonts,amsthm}
\usepackage{graphicx}
\usepackage{booktabs}
\usepackage{array}
\usepackage{braket}
\usepackage{xcolor}
\usepackage[colorlinks=true,linkcolor={blue!50!black},citecolor={blue!50!black},urlcolor={blue!50!black}]{hyperref}

\graphicspath{{figures/}}

\newtheorem{proposition}{Proposition}

\newcommand{\Equiv}{\textsc{Equiv}}
\newcommand{\NonEquiv}{\textsc{NonEquiv}}
\newcommand{\Twirled}{\textsc{NonEquiv-Twirled}}
\newcommand{\NoSym}{\textsc{Equiv-NoSym}}
\newcommand{\tr}{\operatorname{Tr}}
\newcommand{\Ons}{O_{\mathrm{ns}}}
\newcommand{\Hi}{\mathcal{H}}
\newcommand{\RH}{R_{\Hi}}

\begin{document}

\title{Invariance can prevent overfitting and shortcut learning in Quantum Neural Networks}

\author{Giuseppe D'Ambruoso}
\email{g.dambruoso2@phd.poliba.it}
\affiliation{Dipartimento di Meccanica, Matematica e Management, Politecnico di Bari, 70126 Bari, Italy}
\affiliation{Dipartimento Interuniversitario di Fisica, Politecnico di Bari, 70126 Bari, Italy}
\affiliation{Istituto Nazionale di Fisica Nucleare (INFN), Sezione di Bari, 70126 Bari, Italy}

\author{Giovanni Gramegna}
\affiliation{Dipartimento Interuniversitario di Fisica, Universit\`a degli Studi di Bari Aldo Moro, 70126 Bari, Italy}
\affiliation{Istituto Nazionale di Fisica Nucleare (INFN), Sezione di Bari, 70126 Bari, Italy}

\author{Domenico Pomarico}
\affiliation{Dipartimento Interuniversitario di Fisica, Universit\`a degli Studi di Bari Aldo Moro, 70126 Bari, Italy}
\affiliation{Istituto Nazionale di Fisica Nucleare (INFN), Sezione di Bari, 70126 Bari, Italy}

\author{Giuseppe Magnifico}
\affiliation{Dipartimento Interuniversitario di Fisica, Universit\`a degli Studi di Bari Aldo Moro, 70126 Bari, Italy}
\affiliation{Istituto Nazionale di Fisica Nucleare (INFN), Sezione di Bari, 70126 Bari, Italy}

\author{Cosmo Lupo}
\affiliation{Dipartimento Interuniversitario di Fisica, Politecnico di Bari, 70126 Bari, Italy}
\affiliation{Istituto Nazionale di Fisica Nucleare (INFN), Sezione di Bari, 70126 Bari, Italy}

\begin{abstract}
Many image classification tasks have labels that are unchanged by rotations and reflections. Equivariant quantum neural networks incorporate such symmetries by design, but the specific benefit of symmetry is difficult to isolate. Equivariant quantum circuits are usually compared with architectures that also differ in gates, structure or size. Here, we investigate the role of invariance in a controlled setting, classifying images whose labels are invariant under the dihedral group $D_4$. We compare four eight-qubit models with identical encoding, circuit structure and number of trainable parameters, and equivalent dynamical Lie algebras, and benchmark invariant implementations against non-invariant ones in terms of generalization error, robustness to shortcut learning and resilience to a realistic noise model. On four binary tasks, including MNIST and two scientific data sets of galaxies and red blood cells, the invariant models trained on a few tens of images show better generalization than non-invariant ones. Invariance also makes the models immune to orientation shortcuts: when trained with a class-dependent corner watermark, the non-invariant models learn the watermark and often drop to chance level when it is moved. We show that the gain in accuracy of invariant models persists also under a realistic noise model.
\end{abstract}

\maketitle

\section{Introduction}\label{sec:intro}

Variational quantum algorithms, in which a parametrized quantum circuit is trained by a classical optimizer, are a leading approach to machine learning on near-term quantum hardware \cite{preskill2018quantum, cerezo2021variational, schuld2021machine}. Used as classifiers, these quantum neural networks (QNNs) encode classical data into quantum states, process them with a trainable unitary and read out the prediction from the expectation value of an observable \cite{schuld2021machine, havlicek2019supervised, schuld2021effect, larose2020robust}. Two constraints limit their use. The size and noise of current devices limit the number of qubits, parameters and layers, and expressive circuits suffer from barren plateaus, in which the gradients vanish exponentially with the number of qubits \cite{mcclean2018barren, ragone2024lie, fontana2024characterizing, larocca2025barren, wang2021noise}. As a result, QNNs are small and typically trained on small data sets, and whether they generalize from few examples is a central question for their use \cite{caro2022generalization, banchi2021generalization, abbas2021power, gilfuster2024understanding, bowles2024better}.

Symmetry offers a natural way to tailor a model to its task. In classical deep learning, architectures that respect the symmetries of the data yield models that are both expressive and data-efficient: convolutional networks are equivariant under translations, and group-equivariant networks extend this to rotations and reflections \cite{cohen2016group, bronstein2021geometric}. Geometric quantum machine learning transfers this principle to QNNs \cite{cong2019quantum, larocca2022group, meyer2023exploiting, skolik2023equivariant, zheng2023speeding, sauvage2024building, schatzki2024theoretical, ragone2022representation, nguyen2024theory, das2025permutation}: if the circuit is equivariant with respect to a representation of the symmetry group and the measured observable is equivariant, the model output is invariant by construction. Two benefits are expected. First, restricting the model to invariant functions should improve generalization from few examples \cite{kubler2021inductive, elesedy2021provably}. Second, restricting the generators of the circuit reduces its dynamical Lie algebra, which should mitigate barren plateaus \cite{larocca2022group, ragone2024lie, sauvage2024building, west2024provably}. Both have been proven for permutation-equivariant QNNs \cite{schatzki2024theoretical}, but for a small finite group, such as the symmetries of an image, the constraint is weaker, and whether these benefits appear in practice remains an open question. Here we address this question on image classification tasks, including real-world data sets, and find that invariance prevents overfitting and shortcut learning.

In image classification, the relevant geometric symmetries are the transformations under which the image-label association is preserved. For square images, reflections and rotations by multiples of $90^\circ$ are described by the dihedral group $D_4$. Equivariant and approximately equivariant QNNs for this group and its subgroups have been proposed and shown to outperform non-equivariant models on image tasks \cite{chang2023equivariant, west2023reflection, west2024provably, PhysRevResearch.7.013082, das2024role}. In these works, however, the equivariant model is usually compared with generic circuits that differ in their gates, entangling pattern or number of parameters. Ref.~\cite{west2023reflection} partly addresses this gap with a hybrid model that shares the variational circuit of the equivariant one but uses a different encoding, so that the symmetry of the circuit no longer corresponds to a transformation of the images. A systematic comparison on several data sets, in which the circuit structure is fixed and the symmetry is varied in a controlled way, is still missing, and the effect of the symmetry itself remains unclear.
Such a comparison must take into account two facts. First, circuits that differ in their gates also differ in properties that depend on their dynamical Lie algebra (DLA). These include the unitaries they can reach, the number of parameters beyond which they are overparametrized \cite{larocca2023theory} and the size of their gradients \cite{ragone2024lie, fontana2024characterizing}: a difference in generalization may come from any of them. Second, an equivariant model combines two distinct ingredients: an equivariant circuit and an equivariant readout. Only together do they make the model invariant. An invariant model can also be obtained without an equivariant circuit, by averaging a generic model over the group (twirling) \cite{meyer2023exploiting, nguyen2024theory}. In this work we ask which of these ingredients produces the benefit.

\begin{figure*}[tbp]
\centering
\begin{minipage}[c]{0.23\textwidth}\centering
\textbf{a}\\[2pt]
\includegraphics[width=\linewidth]{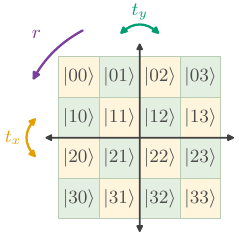}
\end{minipage}\hfill
\begin{minipage}[c]{0.73\textwidth}\centering
\textbf{b}\\[2pt]
\includegraphics[width=\linewidth]{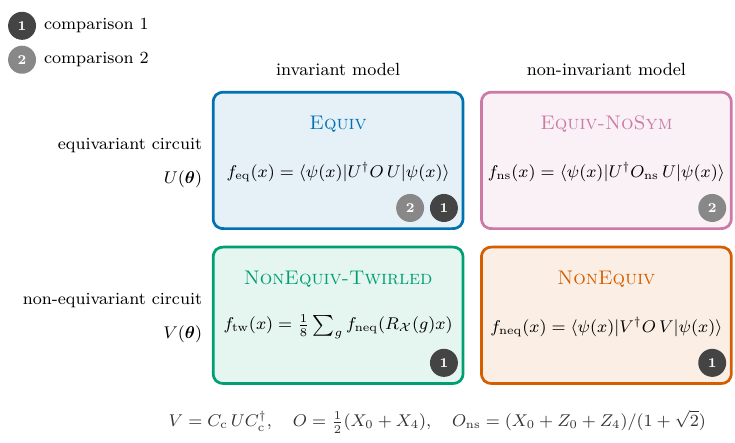}
\end{minipage}
\caption{The study design. \textbf{a} Symmetry-aware amplitude encoding, shown for a $4\times4$ image: pixel $(i,j)$ is the amplitude of the basis state $\ket{i}_{\mathrm r}\otimes\ket{j}_{\mathrm c}$, and the reflections $t_x$, $t_y$ and the rotation $r$, which generate $D_4$, act on the qubits as $X$ layers on one register and a swap of the two registers (Eq.~\eqref{eq:representation}). \textbf{b} The four models. Two circuits, equivariant ($U$) or not ($V$), are combined with an equivariant observable $O$, with a non-equivariant observable $\Ons$, or with the average over $D_4$; the four models have the same circuit structure and trainable angles, and equivalent dynamical Lie algebras. The tags mark the models of the two comparisons of Sec.~\ref{sec:results}: comparison~1 (how the invariance is obtained) and comparison~2 (the same circuit with and without an equivariant readout).}\label{fig:design}
\end{figure*}

We isolate the role of the symmetry with four models that share the same circuit structure (Fig.~\ref{fig:design}). Images are amplitude-encoded so that the group $D_4$ acts on the qubits by bit flips and by a swap of two registers. On this encoding we use an equivariant circuit \citep{chang2023equivariant} and a non-equivariant one that is the same circuit in a rotated basis of one register. The two circuits therefore have the same number and arrangement of trainable angles, and conjugate DLAs of the same dimension. The equivariant circuit measured with an equivariant observable gives an invariant model (\Equiv); measured with a non-equivariant observable, it gives a non-invariant one (\NoSym); the non-equivariant circuit gives a non-invariant model even with the equivariant observable (\NonEquiv), and becomes invariant when averaged over the group (\Twirled). The four models thus form a two-by-two design, in which the equivariance of the circuit and the invariance of the model are varied independently.

We study the regime in which the symmetry is a property of the data: the label is invariant under $D_4$ and the images appear in every pose. We use four such tasks, which differ in origin, image statistics and size: MNIST digits 4 and 5 in random pose \citep{lecun1998gradient}; a synthetic task, a thin cross against a filled square; and two scientific data sets, face-on against edge-on disk galaxies from Galaxy10 \citep{leung2019deep} and sickle-shaped against other deformed red blood cells \citep{gonzalez2015red}. We also test whether the models learn shortcuts, spurious cues that predict the label only in the training data \citep{geirhos2020shortcut}, and whether the results survive the noise of current hardware.

Our main findings are the following.

\begin{enumerate}
\item \emph{Invariance can prevent overfitting, on every task.} With a few tens of training images, the non-invariant models fit the particular images and poses they have seen. The invariant models are also more accurate on new images (Sec.~\ref{subsec:results_c1}).
\item \emph{Circuit equivariance is not enough.} The equivariant circuit measured with a non-equivariant observable (\NoSym) overfits as much as the non-equivariant circuit. What matters is whether the model as a whole is invariant, not whether its circuit is equivariant (Sec.~\ref{subsec:results_c2}).
\item \emph{Overfitting is the cost of learning the symmetry from the data.} On upright digits the same non-equivariant circuit overfits less than half as much, because it no longer has to learn the symmetry from the images (Sec.~\ref{subsec:results_c1}).
\item \emph{Built-in invariance and invariance imposed by classical post-processing (group twirling) both prevent overfitting, but differ in accuracy.} Depending on the task, either can be the more accurate (Sec.~\ref{subsec:results_c1}).
\item \emph{Invariant models are immune to orientation-dependent shortcuts.} Trained with a watermark that reveals the class, the non-invariant models learn the watermark and drop close to chance level without it on the three tasks with real images. The invariant models learn the content of the images (Sec.~\ref{subsec:results_cues}).
\item \emph{The advantage survives a realistic noise model.} It persists under the noise of a typical superconducting quantum processor, whether the models are trained on the noisy device or only run on it (Sec.~\ref{subsec:results_noise}).

\end{enumerate}

The paper is structured as follows. Sec.~\ref{sec:background} recalls equivariance, twirling, DLA and the function class of a model, and shows why an invariant model can learn from fewer images. Sec.~\ref{sec:models} introduces the encoding, the circuits and the four models under consideration, and compares their DLAs and function classes. Sec.~\ref{sec:setup} describes the tasks, the shortcuts, the training and the noise model. Sec.~\ref{sec:results} reports the results and Sec.~\ref{sec:discussion} discusses them. In Sec.~\ref{sec:conclusion}, we draw our conclusions. Technical details, the computation of the DLAs and function classes, and additional results are in the Appendices.
\section{Background}\label{sec:background}

\subsection{Variational quantum classifiers}
We consider an \(n\)-qubit Hilbert space
\(\Hi=(\mathbb{C}^{2})^{\otimes n}\), and denote by
\(\mathcal{B}(\Hi)\) the space of bounded linear operators on
\(\Hi\) and by \(\mathcal{U}(\Hi)\) the subset of unitary operators. A variational quantum model is specified by a family of
parametrized unitaries
\begin{equation}
    U(\boldsymbol{\theta})\in\mathcal{U}(\Hi),
    \qquad
    \boldsymbol{\theta}=(\theta_1,\ldots,\theta_p)\in\mathbb{R}^{p},
\end{equation}
where \(p\) is the number of trainable parameters. Given a classical input
\(x\in\mathcal{X}\), an encoding map associates it with a quantum state
\(\rho(x)\in\mathcal{B}(\Hi)\). The state is then processed by the
variational circuit and measured through an observable
\(O\in\mathcal{B}(\Hi)\), which we call the \emph{readout}, yielding the model output
\begin{equation}
    f_{\boldsymbol{\theta}}(x)
    =
    \tr\!\left[
        O\,U(\boldsymbol{\theta})\rho(x)
        U(\boldsymbol{\theta})^{\dagger}
    \right].
\end{equation}
The encoding, the parametrized circuit, and the readout observable jointly
determine the class of functions that the model can represent.

We now specialize this framework to supervised classification. A model is
trained on a set of labeled examples
\begin{equation}
    \mathcal D=\{(x_i,y_i)\}_{i=1}^{N},
\end{equation}
where \(x_i\in\mathcal{X}\) is the \(i\)-th input data point of the training
set, \(\mathcal{X}\) is the data space, and \(y_i\) is its associated label.
We focus on binary classification tasks, so that \(y_i\in\{0,1\}\), and the model predicts class $1$ if \(f_{\boldsymbol{\theta}}(x)>0\) and \(0\) otherwise. The
examples in \(\mathcal D\) are used to optimize the parameters
\(\boldsymbol{\theta}\) so that the resulting model can predict the labels of
previously unseen inputs.

Training consists of adjusting \(\boldsymbol{\theta}\) with a classical
optimizer to minimize the empirical loss
\begin{equation}\label{eq:Loss}
    \mathcal{L}(\boldsymbol{\theta})
    =
    \frac{1}{N}
    \sum_{i=1}^{N}
    \ell\!\left(f_{\boldsymbol{\theta}}(x_i),y_i\right),
\end{equation}
where \(\ell\) penalizes discrepancies between the model output and the
corresponding labels.

More generally, noisy dynamics can be incorporated by replacing the unitary
evolution with a parametrized quantum channel
\(\Phi_{\boldsymbol{\theta}}:\mathcal{B}(\Hi)\rightarrow
\mathcal{B}(\Hi)\), in which case
\begin{equation}
    f_{\boldsymbol{\theta}}(x)
    =
    \tr\!\left[
        O\,\Phi_{\boldsymbol{\theta}}(\rho(x))
    \right].
\end{equation}
This more general formulation will also allow us to distinguish between the
symmetry properties of the quantum evolution and the invariance of the
resulting classifier, as discussed below.
\subsection{Equivariance, invariance and twirling}\label{subsec:equivariance}
Symmetries of a classification task are transformations of the input
that preserve its target label. For images, these may include rotations
and reflections when orientation is irrelevant to the class. 
We describe a chosen family of such transformations by a finite group \(G\), with representation \(R_{\mathcal X}\) on the input space \(\mathcal X\). Thus, each \(g\in G\) acts as
\begin{equation}
x\mapsto R_{\mathcal X}(g)x,
\end{equation}
and the group operation is represented by composition,
\begin{equation}
R_{\mathcal X}(g_1g_2)=R_{\mathcal X}(g_1)R_{\mathcal X}(g_2).
\end{equation}
For example, rotations
by multiples of $90^\circ$ and reflections of a square form the
dihedral group $D_4$. An invariant classifier assigns the same output
to all inputs related by these transformations.

The action of the group of transformations on quantum states is described by
$\rho\mapsto R_{\Hi}(g)\rho R_{\Hi}(g)^\dagger$, where $R_{\Hi} \colon G\to\mathcal{U}(\Hi)$ is a  unitary representation of $G$ on
the Hilbert space $\Hi$. This representation depends on the particular encoding of the classical data into quantum states $x\mapsto \rho(x)$, with a compatibility condition between the representation $R_\mathcal{X}$ on the data and the representation $\RH$ on the quantum states. More precisely, an encoding is said to be $(G,\RH,R_{\mathcal{X}})$-equivariant if 
\begin{equation}
    \rho(R_{\mathcal{X}}(g)x)=\RH(g)\rho(x)\RH(g)^\dagger.
\end{equation}
Informally, equivariance means that the encoding of the transformed image is the same as the transformed quantum state of the original image. Equivariance of a quantum channel can be defined analogously: 
transforming the input before processing gives the same state as
transforming the output afterwards. Formally, a quantum channel $\Phi:\mathcal{B}(\Hi)\rightarrow\mathcal{B}(\Hi)$ is said to be $(G,\RH)$-\emph{equivariant} if
\begin{equation}
\label{eq:equivariant_channel}
    \Phi(\RH(g)\rho \RH(g)^\dagger)=\RH(g)\Phi(\rho)\RH(g)^\dagger
\end{equation}
for every $g\in G$, and for every $\rho\in\mathcal{B}(\Hi)$ \citep{larocca2022group,meyer2023exploiting,nguyen2024theory}.

A linear operator $O\in\mathcal{B}(\Hi)$ is $(G,\RH)$-\emph{equivariant} if 
\begin{equation}\label{eq:equivariant_operator}
    [O,\RH(g)]=0
\end{equation}
for every $g\in G$. In particular, a parametrized quantum circuit $U(\boldsymbol\theta)=\prod_k e^{-i\theta_kH_k/2}$ is $(G,\RH)$-equivariant for every $\boldsymbol\theta$ if every generator $H_k$ is $(G,\RH)$-equivariant. 
A quantum learning model obtained from an equivariant circuit and an equivariant observable is invariant under the action of the transformation group~\cite{larocca2022group, meyer2023exploiting}.

\begin{proposition}\label{prop:invariance}
If $\Phi$ is a $(G,\RH)$-equivariant quantum channel and $O$ is a $(G,\RH)$-equivariant observable, then the model $f(\rho):=\tr[O\,\Phi(\rho)]$ is $(G,\RH)$-invariant:
\begin{equation}
    f(\RH(g)\rho \RH(g)^\dagger)=f(\rho),
\end{equation}
for all $g\in G$ and $\rho\in\mathcal{B}(\Hi)$.
\end{proposition}
\begin{proof}
    This follows immediately using~\eqref{eq:equivariant_channel}, cyclicity of the trace and the fact that~\eqref{eq:equivariant_operator} implies $\RH(g)^\dagger O \RH(g)=O$.
\end{proof}
In this construction both conditions matter: an equivariant circuit measured with a non-equivariant observable is in general not invariant. A necessary and sufficient condition for invariance is more easily stated in the Heisenberg picture. Let $\Phi^\dagger$ be the adjoint of $\Phi$, defined by $\tr[A\,\Phi(\rho)]=\tr[\Phi^\dagger(A)\,\rho]$. The output of a model is $f_{\boldsymbol\theta}(x)=\tr[A_{\boldsymbol\theta}\,\rho(x)]$, with the \emph{effective observable}
\begin{equation}
A_{\boldsymbol\theta}=\Phi_{\boldsymbol\theta}^\dagger(O),
\label{eq:effective_observable}
\end{equation}
which is $U(\boldsymbol\theta)^\dagger O\,U(\boldsymbol\theta)$ for a unitary circuit. Since $\tr[A\,\RH(g)\rho\RH(g)^\dagger]=\tr[\RH(g)^\dagger A\RH(g)\,\rho]$, the model is invariant on all input states if and only if
\begin{equation}
\RH(g)^\dagger A_{\boldsymbol\theta}\,\RH(g)=A_{\boldsymbol\theta}\qquad\text{for every }g\in G.
\label{eq:invariance_criterion}
\end{equation}
Proposition~\ref{prop:invariance} is the special case in which this holds because $\Phi_{\boldsymbol\theta}$ and $O$ are both equivariant. When the encoded states do not span all the states, as for the real images of Sec.~\ref{subsec:encoding}, the condition is required only of the part of $A_{\boldsymbol\theta}$ that they detect (Sec.~\ref{subsec:dla}).

An invariant model can also be obtained from a non-equivariant circuit by averaging its output over the group, an operation called \emph{twirling} \citep{meyer2023exploiting,nguyen2024theory}. For an equivariant encoding,
\begin{equation}
f_{\mathrm{tw}}(x):=\frac{1}{|G|}\sum_{g\in G}f_{\boldsymbol\theta}(R_{\mathcal X}(g)x)=\tr\bigl[\mathcal P_G(A_{\boldsymbol\theta})\,\rho(x)\bigr],
\label{eq:twirl}
\end{equation}
where
\begin{equation}
\mathcal P_G(A):=\frac{1}{|G|}\sum_{g\in G}\RH(g)^\dagger A\,\RH(g)
\label{eq:group_average}
\end{equation}
is the projection onto the operators that commute with the representation. The twirled model therefore satisfies Eq.~\eqref{eq:invariance_criterion} for every channel and every observable; if $O$ is equivariant, it coincides with the model obtained by twirling the channel instead \citep{nguyen2024theory}. On a physical device, Eq.~\eqref{eq:twirl} is estimated by running the circuit on the transformed inputs: all $|G|$ of them, fewer if the model already has part of the symmetry (Sec.~\ref{subsec:models}), or a random sample of them. In classical simulation, $\mathcal P_G(A_{\boldsymbol\theta})$ can instead be computed once per parameter update and reused for all the inputs of a batch.

\subsection{Dynamical Lie algebra and function class}\label{subsec:terms}

A quantum model is built from three ingredients: the gates of the circuit, the readout observable and the encoding. We characterize it at three nested levels, each of which adds one ingredient: the dynamical Lie algebra depends only on the gates, the span of the effective observables also on the readout, and the feature span also on the encoding.

\paragraph{Dynamical Lie algebra.} The dynamical Lie algebra (DLA) of a set of gates $e^{-i\theta_kH_k/2}$ is the real Lie algebra generated by their generators \citep{larocca2023theory,ragone2024lie},
\begin{equation}
\mathfrak g=\mathrm{span}_{\mathbb R}\bigl\{iH_k,\;[iH_k,iH_l],\;[iH_k,[iH_l,iH_m]],\;\ldots\bigr\}.
\label{eq:dla}
\end{equation}
Every circuit built from these gates lies in the Lie group $e^{\mathfrak g}$, for every choice of the angles and every depth, and deep enough circuits with independent angles reach all of it. The DLA thus describes the unitaries available to the circuit. It also determines bounds on the number of parameters needed to overparametrize a circuit \citep{larocca2023theory} and, for deep circuits and under assumptions on the input states and the observable, the variance of the gradients, which is exponentially small when $\mathfrak g$ is exponentially large \citep{ragone2024lie,fontana2024characterizing}; neither quantity is fixed by the dimension of $\mathfrak g$ alone. If every generator commutes with $\RH(g)$, so does every element of $\mathfrak g$: an equivariant circuit has a DLA contained in the commutant of the representation, which is the sense in which symmetry restricts the DLA \citep{larocca2022group}. The DLA, however, sees neither the observable nor the encoding. A fixed change of basis $W$ maps a circuit and its DLA to conjugate ones, which reach the same unitaries up to $W$, but with the same encoding and readout it can change completely the functions that the model computes.

\paragraph{Effective observables.} The effective observables $A_{\boldsymbol\theta}=U(\boldsymbol\theta)^\dagger OU(\boldsymbol\theta)$ of all the circuits built from the gates span the space
\begin{equation}
\mathcal M_O=\mathrm{span}\{U^\dagger OU:\ U\in e^{\mathfrak g}\},
\label{eq:module}
\end{equation}
which is the smallest space that contains $O$ and is closed under commutators with $\mathfrak g$, that is, the span of $O$ and of its nested commutators $[iH_{k_1},[iH_{k_2},\ldots[iH_{k_r},O]]]$ \citep{ragone2024lie,diaz2023showcasing}. Its dimension depends on the gates and on the observable. If the circuit and $O$ are both equivariant, every element of $\mathcal M_O$ is equivariant.

\paragraph{Function class and feature span.} The function class of a model is the family $\mathcal F=\{x\mapsto f_{\boldsymbol\theta}(x)\}$ of the functions that it computes as its angles vary, a nonlinear family with at most as many local degrees of freedom as trainable angles. We characterize it by its linear span, the \emph{feature span}, and call \emph{features} the elements of a basis of this span. The feature span is the image of the span of the effective observables, which is contained in $\mathcal M_O$, under the linear map $A\mapsto\tr[A\,\rho(\cdot)]$, so its dimension is at most $\dim\mathcal M_O$, and smaller when different observables coincide on all the encoded states. This dimension counts the directions that the outputs can explore, not the number of independently adjustable coefficients, and a larger span does not by itself imply better or worse generalization. If every element of $\mathcal M_O$ is equivariant and the encoding is equivariant, every function in $\mathcal F$ is invariant (Eq.~\eqref{eq:invariance_criterion}). Twirling replaces every effective observable with its projection $\mathcal P_G(A_{\boldsymbol\theta})$: the resulting feature span is invariant, but in general different from that of an equivariant circuit.

The three levels answer different questions. The DLA describes the unitaries available to the circuit; the effective observables take into account the readout measurement; the feature span describes the functions available to the model, and its invariance---whether the output can depend on the pose of the input. Two models with the same DLA can differ in the last two (Sec.~\ref{subsec:dla}).

Invariance forces a model to assign the same output to all symmetry-related images, thus preventing the model from exploiting pose-dependent patterns in the finite training sample. When pose carries no information about the target label, this constraint supplies an appropriate inductive bias. However, it can still memorize the training images and their transformed copies without generalizing to new examples. For example, two handwritten ``$4$''s generally belong to different orbits: rotating or reflecting one does not produce the other. An invariant model may correctly classify every transformed version of a training ``$4$'', yet misclassify a new ``$4$'' with different handwriting.

For an invariant model, the training loss on $N$ images equals the loss on all their $N|G|$ transformed copies: the model is trained as if on every pose of its images, although these copies are not independent samples.
\section{Models}\label{sec:models}
\begin{figure*}[t]
\centering
\includegraphics[width=\textwidth]{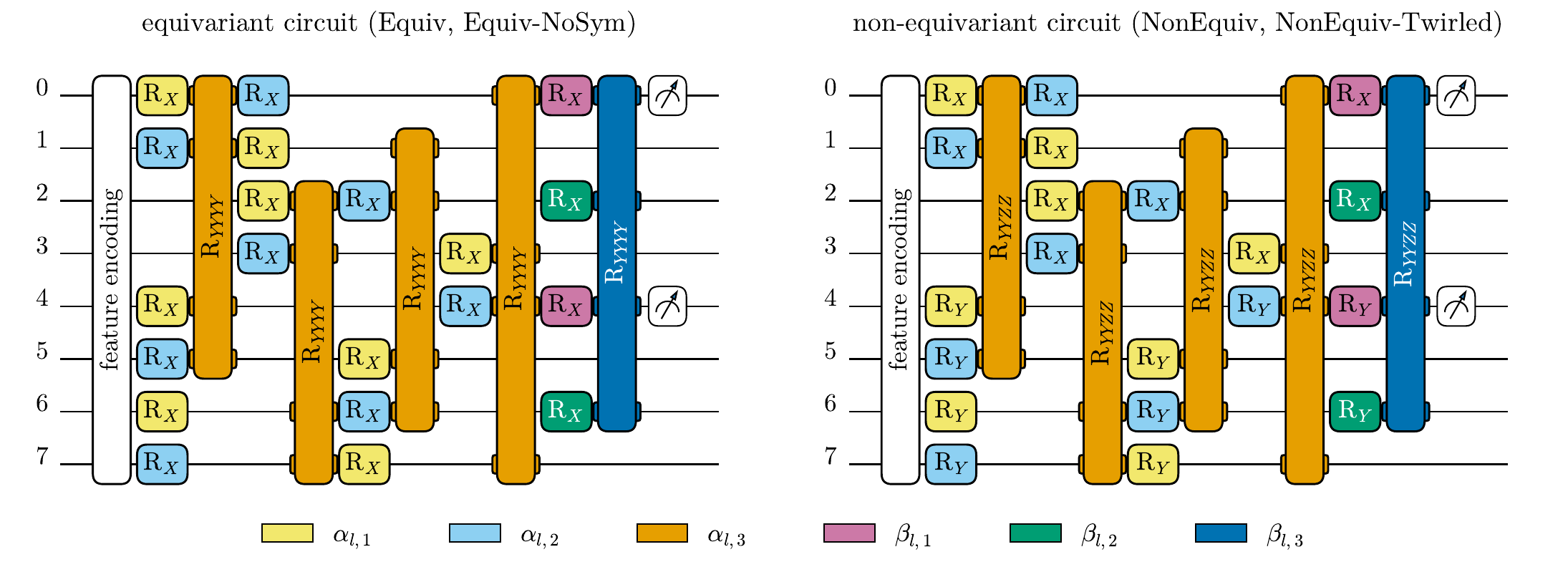}
\caption{One layer of the two circuits described in Sec.~\ref{subsec:circuits} (wires $0$--$3$: row register; wires $4$--$7$: column register). The first box is the feature encoding, which prepares the state $\ket{\psi(x)}$ of Eq.~\eqref{eq:encoding}; $\mathrm R_P(\eta)=e^{-i\eta P/2}$ is the rotation generated by the Pauli string $P$, and gates with the same color share the angle $\eta$, a component of $\boldsymbol\alpha_l$ or $\boldsymbol\beta_l$ (Eq.~\eqref{eq:circuit_layer}). The non-equivariant circuit (right) has $\mathrm R_Y$ rotations on the column register and $\mathrm R_{YYZZ}$ rotations in place of $\mathrm R_X$ and $\mathrm R_{YYYY}$, with the same gate count, wiring and depth.}\label{fig:circuits}
\end{figure*}

\subsection{Symmetry-aware encoding}\label{subsec:encoding}
Each grayscale image is resized to $d\times d$ pixels, where
$d=2^{n/2}$ and $n$ is a positive even integer. Its pixel intensities
are rescaled to real $\ell_2$-normalized coefficients $x_{ij}$ and amplitude-encoded on $n$ qubits
\citep{schuld2021machine,yao2017quantum},
\begin{equation}
\ket{\psi(x)}
=\sum_{i,j=0}^{d-1}x_{ij}\,\ket{i}_{\mathrm r}\otimes\ket{j}_{\mathrm c}.
\label{eq:encoding}
\end{equation}
The row and column indices are stored in binary, with the most
significant bit first, in two registers of $n/2$ qubits each.
The row register occupies wires $0,\ldots,n/2-1$, and the column
register occupies wires $n/2,\ldots,n-1$. The case $n=8$, corresponding
to $16\times16$ pixels, is illustrated in Fig.~\ref{fig:design}a.

The symmetries of the square pixel grid form the dihedral group
\begin{equation}
D_4=\{e,r,r^2,r^3,t,tr,tr^2,tr^3\},
\end{equation}
where $e$ is the identity, $r$ is the anticlockwise rotation by
$90^\circ$, and $t$ is the reflection about the vertical axis.
Thus, the elements $r^k$ describe rotations by multiples of
$90^\circ$, while the elements $tr^k$ describe reflections about
the horizontal, vertical, and two diagonal axes.
The encoding induces the unitary representation
\begin{equation}
\begin{aligned}
R_{\mathcal H}(t_x)
\bigl(\ket{i}_{\mathrm r}\otimes\ket{j}_{\mathrm c}\bigr)
&=\ket{d-1-i}_{\mathrm r}\otimes\ket{j}_{\mathrm c},\\
R_{\mathcal H}(t_y)
\bigl(\ket{i}_{\mathrm r}\otimes\ket{j}_{\mathrm c}\bigr)
&=\ket{i}_{\mathrm r}\otimes\ket{d-1-j}_{\mathrm c}.
\end{aligned}
\end{equation}
Here $t_y=t$ is the reflection about the vertical axis and
$t_x=tr^2$ the reflection about the horizontal axis.
Since $d=2^{n/2}$, replacing an index $i$ by $d-1-i$ complements
all its binary digits. These reflections are therefore implemented
by applying Pauli $X$ gates to every qubit of the corresponding
register.

With rows counted from top to bottom and columns from left to
right, the anticlockwise rotation by $90^\circ$ maps pixel positions
as $(i,j)\mapsto(d-1-j,i)$. Its representation then satisfies
\begin{equation}
R_{\mathcal H}(r)
\bigl(\ket{i}_{\mathrm r}\otimes\ket{j}_{\mathrm c}\bigr)
=\ket{d-1-j}_{\mathrm r}\otimes\ket{i}_{\mathrm c},
\end{equation}
which can be obtained by first swapping the two registers
and then applying $R_{\mathcal H}(t_x)$. The register swap itself
represents the reflection $tr^3$ about the main diagonal.
Consequently,
\begin{equation}
\begin{aligned}
R_{\mathcal H}(t_x)
&=X^{\otimes(n/2)}\otimes\mathbb I^{\otimes(n/2)},\\
R_{\mathcal H}(t_y)
&=\mathbb I^{\otimes(n/2)}\otimes X^{\otimes(n/2)},\\
R_{\mathcal H}(r)
&=R_{\mathcal H}(t_x)\,S,\\
S:=R_{\mathcal H}(tr^3)
&=\prod_{a=0}^{n/2-1}\mathrm{SWAP}_{a,a+n/2}.
\end{aligned}
\label{eq:representation}
\end{equation}
The representations of the remaining group elements follow by
composition. By construction, the encoding is $D_4$-equivariant:
\begin{equation}
\ket{\psi(R_{\mathcal X}(g)x)}
=R_{\mathcal H}(g)\ket{\psi(x)}
\qquad\text{for every }g\in D_4.
\end{equation}

\subsection{Equivariant circuits and controlled symmetry breaking}
\label{subsec:circuits}

To isolate the role of circuit equivariance, we construct two
families with the same parameter count, parameter sharing and
gate connectivity. By \emph{parameter sharing} we mean that a
single trainable angle is used by several gates, which are then
rotated together during training; two circuits have the same
parameter sharing when their gates use the angles in the same
pattern. In both circuits, the rotations of a row qubit and of its
column partner share one angle, as required by the symmetry under
the swap of the two registers, and in every layer the four blocks
on the bit pairs $(0,1)$, $(2,3)$, $(1,2)$ and $(3,0)$ share three
angles, while the block on $(0,2)$ has three of its own
(Eq.~\eqref{eq:circuit_layer}): the $25$ gates of a layer depend on
only six trainable angles. In the code (\texttt{eqnn/circuits.py}),
each gate is stored with the index of the entry of
$\boldsymbol\theta$ that it uses, and the gradient with respect to
a shared angle, the sum of the contributions of all the gates that
use it, is computed by automatic differentiation. We first build a $D_4$-equivariant circuit
$U(\boldsymbol\theta)$, then define a comparison circuit
$V(\boldsymbol\theta)$ by conjugating $U(\boldsymbol\theta)$
with a fixed unitary acting on the column register.
Throughout this construction, the encoding and the representation
of the image symmetries are kept fixed.

\begin{figure*}[tbp]
\centering
\includegraphics[width=\textwidth]{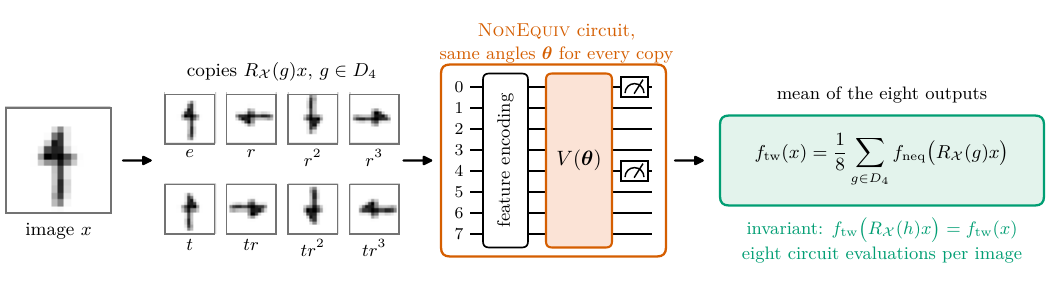}
\caption{\Twirled: the image is transformed by the eight elements of $D_4$, the copies go through the same \NonEquiv\ circuit with the same angles, and the output is the mean of the eight expectation values. The model is exactly invariant, at eight times the cost of \NonEquiv. }\label{fig:twirled}
\end{figure*}

We label the row and column qubits by $\mathrm r_j$ and $\mathrm c_j$,
respectively, with $j=0,\ldots,n/2-1$; these correspond to wires
$j$ and $j+n/2$ in Fig.~\ref{fig:circuits}.
For two bit positions $j,k$, define the four-qubit unitary
\begin{equation}
\begin{aligned}
W_{jk}(\boldsymbol\eta)
={}&e^{-i\frac{\eta_3}{2}
Y_{j}Y_{k}Y_{j+n/2}Y_{k+n/2}}
\\
&\times e^{-i\frac{\eta_2}{2}(X_{k}+X_{k+n/2})}
e^{-i\frac{\eta_1}{2}(X_{j}+X_{j+n/2})},
\end{aligned}
\label{eq:equiv_block}
\end{equation}
where $\boldsymbol\eta=(\eta_1,\eta_2,\eta_3)$.
Each rotation acts in the same way on a row qubit and on its
column partner, and the four-qubit rotation acts on both
registers with the same Pauli operators. This makes the block
symmetric under the reflections of either register and under
their swap, and hence $D_4$-equivariant for every choice of the
angles.

The circuit follows the arrangement in Fig.~\ref{fig:circuits},
adapted from \citet{chang2023equivariant}.
In layer $l$, the four pairs
$(0,1)$, $(2,3)$, $(1,2)$ and $(3,0)$ share a parameter triple
$\boldsymbol\alpha_l$, while the pair $(0,2)$ uses an
independent triple $\boldsymbol\beta_l$.
With the rightmost factor acting first, the layer is
\begin{equation}
\begin{aligned}
U_l
={}&W_{02}(\boldsymbol\beta_l)
W_{30}(\boldsymbol\alpha_l)
W_{12}(\boldsymbol\alpha_l)
 W_{23}(\boldsymbol\alpha_l)
W_{01}(\boldsymbol\alpha_l),
\end{aligned}
\label{eq:circuit_layer}
\end{equation}
and the full circuit is
\begin{equation}
U(\boldsymbol\theta)=U_L\cdots U_1,
\qquad
\boldsymbol\theta=
(\boldsymbol\alpha_1,\boldsymbol\beta_1,\ldots,
 \boldsymbol\alpha_L,\boldsymbol\beta_L).
\label{eq:full_equivariant_circuit}
\end{equation}
Thus, an $L$-layer circuit has $6L$ trainable parameters and
is $D_4$-equivariant for every choice of these parameters.

To construct the comparison circuit, let 
\begin{equation}
    C=\exp\left({-i\frac{\pi}{3}\frac{X+Y+Z}{\sqrt{3}}}\right)
\end{equation}
be the
single-qubit Clifford unitary that cycles the Pauli axes,
\begin{equation}
    CXC^\dagger=Y,\qquad
CYC^\dagger=Z,\qquad
CZC^\dagger=X,
\end{equation}
and set $C_{\mathrm c}=\mathbb I^{\otimes(n/2)}\otimes C^{\otimes(n/2)}$.
We define
\begin{equation}
V(\boldsymbol\theta)
:=C_{\mathrm c}\,U(\boldsymbol\theta)\,C_{\mathrm c}^\dagger.
\label{eq:conjugate_circuits}
\end{equation}
Conjugation acts on each block as
\begin{equation}
\begin{aligned}
C_{\mathrm c}W_{jk}(\boldsymbol\eta)C_{\mathrm c}^\dagger
={}&e^{-i\frac{\eta_3}{2}
Y_{j}Y_{k}Z_{j+n/2}Z_{k+n/2}}
\\
&\times e^{-i\frac{\eta_2}{2}(X_{k}+Y_{k+n/2})}
e^{-i\frac{\eta_1}{2}(X_{j}+Y_{j+n/2})}.
\end{aligned}
\label{eq:nonequiv_block}
\end{equation}
Implementing these conjugated gates directly preserves the
gate count, connectivity, depth and parameter sharing.

The transformed generators no longer all commute with the
column reflection or with the swap of the two registers.
Consequently, $V(\boldsymbol\theta)$ is generally not
$D_4$-equivariant with respect to the fixed representation
\eqref{eq:representation}. It nevertheless retains equivariance
under the row reflection $t_x$, since this reflection commutes
with both $U(\boldsymbol\theta)$ and $C_{\mathrm c}$.

Finally, unitary conjugation preserves commutators, so the
dynamical Lie algebras of the two circuit families satisfy
\begin{equation}
\mathfrak g_V
=C_{\mathrm c}\,\mathfrak g_U\,C_{\mathrm c}^\dagger,
\qquad
\dim\mathfrak g_V=\dim\mathfrak g_U.
\label{eq:conjugate_dla}
\end{equation}

\subsection{Four models}\label{subsec:models}

The output of a model is an expectation value, with no trainable classical post-processing: the predicted class is $1$ if $f(x)>0$ and $0$ otherwise. We measure either
\begin{equation}
O=\tfrac12\left(X_0+X_4\right)\quad\text{or}\quad \Ons=\frac{X_0+Z_0+Z_4}{1+\sqrt2},
\label{eq:observables}
\end{equation}
both normalized so that $f(x)\in[-1,1]$. $O$ is equivariant, while $\Ons$ breaks every symmetry of $D_4$ ($Z_0$ and $Z_4$ change sign under the reflections, and the swap maps $X_0+Z_0$ to $X_4+Z_4$). The four models of Fig.~\ref{fig:design}b, \Equiv\ ($f_{\mathrm{eq}}$), \NonEquiv\ ($f_{\mathrm{neq}}$), \Twirled\ ($f_{\mathrm{tw}}$) and \NoSym\ ($f_{\mathrm{ns}}$), are
\begin{equation}
\begin{split}
f_{\mathrm{eq}}(x)&=\bra{\psi(x)}U^\dagger O\,U\ket{\psi(x)},\\
f_{\mathrm{neq}}(x)&=\bra{\psi(x)}V^\dagger O\,V\ket{\psi(x)},\\
f_{\mathrm{tw}}(x)&=\tfrac18\textstyle\sum_{g\in D_4}f_{\mathrm{neq}}(R_\mathcal{X}(g)x),\\
f_{\mathrm{ns}}(x)&=\bra{\psi(x)}U^\dagger \Ons\,U\ket{\psi(x)}.
\end{split}
\label{eq:models}
\end{equation}
\Equiv\ is invariant by Proposition~\ref{prop:invariance} and \Twirled\ by construction: the eight transformed copies of the image, obtained classically by permuting the pixels, go through the same circuit with the same angles (Fig.~\ref{fig:twirled}).

\NonEquiv\ and \NoSym\ are not invariant under $D_4$, and they differ in how the invariance is broken. 
\NonEquiv\ retains invariance under the row reflection $t_x$, since
both its circuit $V$ and its readout observable $O$ commute with
$R_{\mathcal H}(t_x)$. It is therefore invariant under the subgroup
$\{e,t_x\}$: for any $g\in D_4$, applying $t_x$ after $g$ leaves the model output
unchanged, the inputs $R_{\mathcal X}(g)x$ and
$R_{\mathcal X}(t_xg)x$ giving identical outputs.
Since $t_x^2=e$, this groups the eight transformations into four
disjoint pairs $\{g,t_xg\}$, called the \emph{right cosets} of
$\{e,t_x\}$ in $D_4$. Consequently, the eight terms defining
$f_{\mathrm{tw}}$ form four pairs of equal terms in the noiseless
model, and only one evaluation per pair is needed.
The noise model of Sec.~\ref{subsec:noise} breaks this residual
symmetry and removes the reduction.

\NoSym\ breaks invariance through its readout: although its circuit
is equivariant, the observable $\Ons$ breaks every nontrivial
symmetry of $D_4$. Thus, this model has no residual symmetry
enforced by its architecture.

The relation between \NoSym\ and \Equiv\ can be characterized explicitly. Writing $\Ons=(O+Q)/(1+\sqrt2)$, with $Q=\tfrac12(X_0-X_4)+Z_0+Z_4$ and $\mathcal P_G(Q)=0$,
\begin{equation}
f_{\mathrm{ns}}(x)=\frac{f_{\mathrm{eq}}(x)+h_{\boldsymbol\theta}(x)}{1+\sqrt2},
\label{eq:nosym_decomposition}
\end{equation}
where $h_{\boldsymbol\theta}(x)=\bra{\psi(x)}U^\dagger Q\,U\ket{\psi(x)}$ averages to zero over the group because $U$ commutes with every $\RH(g)$. \NoSym\ thus adds a pose-dependent term to a rescaled \Equiv; the same angles set both terms, which cannot in general be tuned independently. Twirling \NoSym\ removes this term and returns \Equiv, up to the constant factor $1/(1+\sqrt2)$.

\subsection{Same algebra, different function classes}\label{subsec:dla}

The two parametrized circuits $U(\bm{\theta})$ and $V(\bm{\theta})$ have conjugate DLAs, and \NoSym\ and \Twirled\ inherit the circuits of \Equiv\ and \NonEquiv. All four models therefore share the DLA dimension, $\dim\mathfrak g=569$, which we computed with the Lie closure of PennyLane (\texttt{qml.lie\_closure}) \citep{bergholm2018pennylane} (Appendix~\ref{app:dla}). The four models are thus equivalent in what the DLA alone determines, the reachable unitaries (up to the fixed change of basis $C_{\mathrm c}$) and the overparametrization threshold: no circuit is larger or more expressive than another. Their gradients depend also on the observable and on the inputs, so we measured them in the regime of our training (Fig.~\ref{fig:gradients} in Appendix~\ref{app:gradients}): none of them vanishes, and they differ by at most a factor of a few, \NoSym\ having the smallest. All four models reach comparable training accuracies (Fig.~\ref{fig:main}a--d), along similar learning curves (Fig.~\ref{fig:curves}).

\begin{figure}[tbp]
\centering
\includegraphics[width=\linewidth]{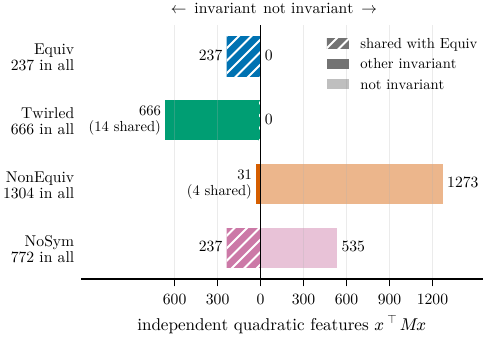}
\caption{Equivalent DLAs, different function classes: number of independent quadratic features $x^{\mathsf T}Mx$ of the image combined by each model, split into $D_4$-invariant (left) and non-invariant (right) ones; hatched: features shared with \Equiv. All four models have a DLA of dimension $569$ (Appendix~\ref{app:dla}).}\label{fig:function_classes}
\end{figure}

What differs is whether the symmetry of the circuit coincides with the action of $D_4$ on the images and with the symmetry of the readout, and this fixes the function class (Fig.~\ref{fig:function_classes}). \Equiv\ combines $237$ features, all invariant. \NonEquiv\ combines $1304$, almost none of which is invariant. \Twirled\ projects them onto $666$ invariant features, nearly disjoint from those of \Equiv. \NoSym\ combines the $237$ features of \Equiv\ and $535$ non-invariant ones. The invariance of a model, which the DLA does not see, decides whether its features can depend on the pose of the image.

\begin{figure*}[tbp]
\centering
\includegraphics[width=\textwidth]{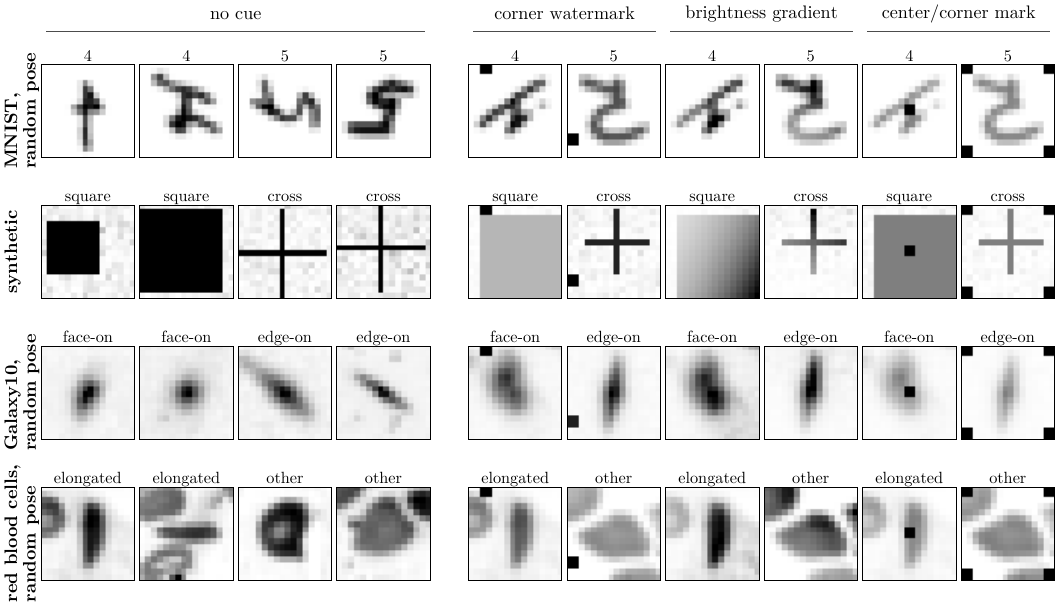}
\caption{The four tasks (rows). Left: training images without cues. Right: one image of each class with each of the cues used in the experiments on shortcuts (Sec.~\ref{subsec:shortcuts}). The corner watermarks and the brightness gradients of the two classes are related by a rotation of $90^\circ$; the center/corner mark is invariant under $D_4$.}\label{fig:tasks}
\end{figure*}

\section{Setup}\label{sec:setup}

\subsection{Tasks}\label{subsec:tasks}

We use four binary tasks (Fig.~\ref{fig:tasks}): MNIST digits $4$ against $5$ \citep{lecun1998gradient}; a synthetic task, a thin cross against a filled square, of random size and position on a weak noisy background; face-on against edge-on disk galaxies from Galaxy10 SDSS \citep{leung2019deep,lintott2008galaxy}; and elongated (sickle) against other deformed red blood cells from erythrocytesIDB \citep{gonzalez2015red,erythrocytesidb}. In all of them the label is invariant under $D_4$, and every image, in training and in test, appears in a random pose: MNIST, Galaxy10 and the blood cells are transformed by an element of $D_4$ drawn at random, and the synthetic shapes are themselves symmetric and placed at random. On the galaxies and the blood cells the orientation is already random by nature, and the random pose only makes the invariance exact. As a control we also train and test on upright MNIST digits. The training sets are balanced and drawn, for each seed, from a pool disjoint from the test set. Appendix~\ref{app:tasks} describes the data sets and how the synthetic task and the pairs of classes were chosen.

\subsection{Shortcuts}\label{subsec:shortcuts}

To test what the models learn when the training set offers an easier rule than the true one \citep{geirhos2020shortcut}, we add to every training image a class-dependent cue (Fig.~\ref{fig:tasks}, right): a \emph{corner watermark}, a small patch in one corner for class $0$ and in the corner rotated by $90^\circ$ for class $1$; a \emph{brightness gradient} along an oblique direction, rotated by $90^\circ$ for class $1$; and, as a control, a \emph{center/corner mark}, a bright patch at the center for class $0$ and in the four corners for class $1$. The watermarks and the gradients of the two classes are related by an element of $D_4$, and the content of the images is in a random pose, so an invariant model cannot tell them apart; the control cue is itself invariant, and every model can read it. Each model is tested on images with the cue of their class, as in training; with the cued image moved by a random element of $D_4$, which in general carries the cue to a position not seen in training, or onto the cue of the other class; and without the cue, which measures whether the model has learned the true rule.

\begin{figure*}[tbp]
\centering
\includegraphics[width=\textwidth]{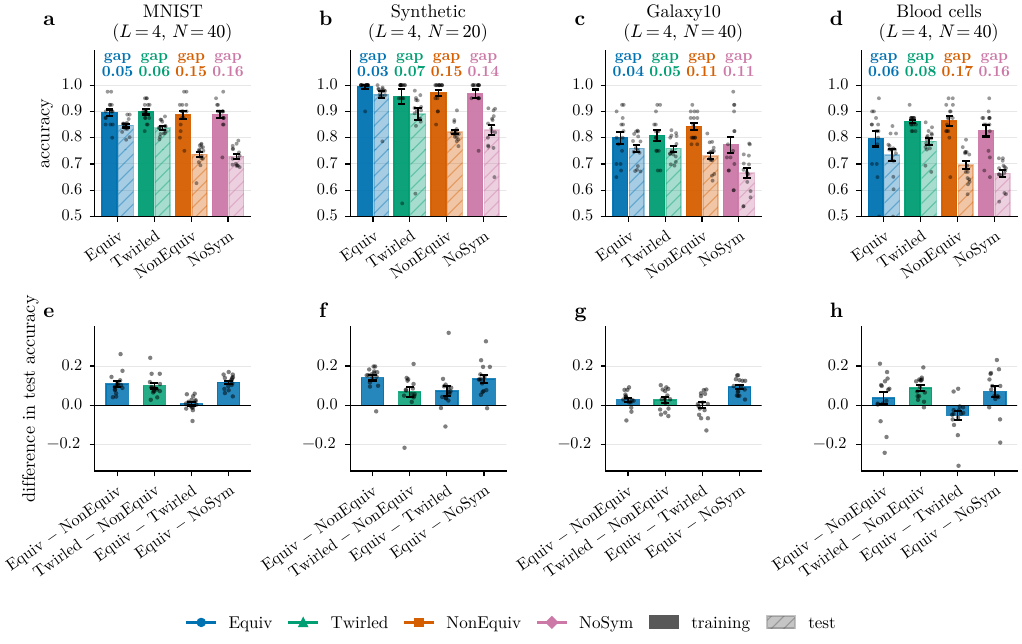}
\caption{The four models in the main settings ($L=4$; $N=40$, or $20$ on the synthetic task; $15$ seeds). \textbf{a--d} Training (solid) and test accuracy (hatched), mean and standard error, with the single seeds (dots) and the generalization gap of each model above its bars. \textbf{e--h} Per-seed differences in test accuracy between pairs of models. The invariant models (\Equiv, \Twirled) have small gaps on every task, the non-invariant ones (\NonEquiv, \NoSym) large ones. Paired comparisons in Table~\ref{tab:paired}; values in the repository.}\label{fig:main}
\end{figure*}

\subsection{Training and evaluation}\label{subsec:training}

The circuits are simulated exactly, with expectation values computed without shot noise, and trained with Adam on a logistic loss of the output; all the details are in Appendix~\ref{app:training}. We report the training accuracy, the test accuracy and their difference, the \emph{generalization gap} \cite{Pomarico2025}. A seed fixes the training images, the poses, the initial angles and the mini-batches, and every model is trained with the same seeds, so all the comparisons are paired; differences are tested with the Wilcoxon signed-rank test. The results in the main text are for the \emph{main settings}, $L=4$ layers ($24$ angles) and $N=40$ training images ($N=20$ on the synthetic task), with $15$ seeds; this is the depth at which the non-equivariant circuit can fit its training images and the equivariant one trains reliably (Appendix~\ref{app:training}). The dependence on $N$ and $L$ is in Appendix~\ref{app:scaling}, and every other setting in the repository of the paper.

\subsection{Hardware noise}\label{subsec:noise}

We also simulate the circuits under a noise model with the typical parameters of current superconducting processors: after every gate, thermal relaxation and depolarizing noise on each of its qubits, relaxation of the idle qubits, and asymmetric readout errors. All the error channels act on single qubits, including those of the multi-qubit gates, with no correlated errors, and their parameters differ from qubit to qubit (Appendix~\ref{app:noise}). A global scale $s$ multiplies all the error rates: $s=0$ is the noiseless circuit and $s=1$ a typical device, with the median parameters of current superconducting processors (Table~\ref{tab:noise_params}). We consider two scenarios in the main settings: models trained and tested under noise ($s$ up to $4$, five seeds), and models trained without noise and only run on the noisy device ($15$ seeds). Realistic noise breaks the symmetry of \Equiv, because amplitude damping does not commute with the reflections and the qubits exchanged by the swap have different parameters, but only slightly: on a typical device its output changes by less than $2\%$ under the group (Appendix~\ref{app:noise}). \Twirled, whose symmetry is classical, stays exactly invariant.
\section{Results}\label{sec:results}

Figure~\ref{fig:main} summarizes the results without cues and without noise. The four tasks differ in the origin and statistics of the images and in how hard they are for the circuits, yet the picture is the same on all of them: the two invariant models do not overfit, and the two non-invariant ones do.

\subsection{Comparison 1: invariance improves generalization}\label{subsec:results_c1}

The four models fit their training images comparably, but generalize very differently (Fig.~\ref{fig:main}a--d). On every task the generalization gap of \NonEquiv\ is two to five times that of the invariant models, and the difference holds in most seeds (Table~\ref{tab:paired}). The invariant models are also more accurate on new images: \Equiv\ beats \NonEquiv\ in at least $12$ of the $15$ seeds on every task (Fig.~\ref{fig:main}e--h, Table~\ref{tab:paired}). The learning curves show how the gap forms: after a few epochs the test accuracy of \NonEquiv\ stops growing while its training accuracy keeps rising, as the circuit fits the particular images and poses of the training set; for the invariant models the two grow together (Fig.~\ref{fig:curves}). The advantage is largest with few images and shrinks as $N$ grows, because more images show more poses (Appendix~\ref{app:scaling}).

\paragraph{The role of the pose.} The control on upright digits identifies the cause. On upright images the same \NonEquiv\ circuit overfits less than half as much as in random pose, and it is even more accurate than \Equiv, because a fixed orientation is informative and a non-invariant model can use it. The advantage is fragile, though: on rotated digits the same trained circuit drops to $0.64$ (not shown; values in the repository). The overfitting of \NonEquiv\ in random pose is thus the cost of learning the symmetry from the data, which the invariant models do not pay.

\paragraph{Built-in or averaged invariance.} Both invariant models avoid overfitting, but which one is more accurate depends on the task. On MNIST and Galaxy10 they are equivalent. On the synthetic task \Equiv\ is clearly better, and \Twirled\ approaches it only with more images; on the blood cells \Twirled\ is better, because \Equiv\ does not fit its training images as well. The function classes explain both cases. That of \Twirled\ is almost three times as large and nearly disjoint from that of \Equiv\ (Sec.~\ref{subsec:dla}). A small class generalizes from fewer images when it contains a good rule, as on the synthetic task; when it does not, as on the blood cells, a larger class fits the images better without overfitting more.

\subsection{Comparison 2: the equivariant circuit needs an equivariant readout}\label{subsec:results_c2}

\begin{table*}[tbp]
\footnotesize
\begin{tabular*}{\textwidth}{@{\extracolsep{\fill}}l cc cc@{}}
\toprule
Task & test acc. & $p$ & gap & $p$\\
\midrule
MNIST & $+0.11(1)$ & $6\times10^{-5}$ & $+0.10(1)$ & $6\times10^{-5}$ \\
Synthetic & $+0.10(2)$ & $2\times10^{-4}$ & $+0.09(1)$ & $6\times10^{-5}$ \\
Galaxy10 & $+0.06(1)$ & $1\times10^{-4}$ & $+0.07(1)$ & $6\times10^{-5}$ \\
Blood cells & $+0.08(2)$ & $3\times10^{-3}$ & $+0.10(2)$ & $6\times10^{-5}$ \\
\bottomrule
\end{tabular*}
\caption{Effect of the invariance of the model in the main settings ($15$ seeds): per seed, the mean of the invariant models (\Equiv, \Twirled) minus the mean of the non-invariant ones (\NoSym, \NonEquiv). For the test accuracy a positive effect is an increase, for the gap a reduction; mean with the standard error on the last digit, and Wilcoxon $p$ value.}\label{tab:factorial}
\end{table*}

Comparison~1 does not separate the circuit from the invariance of the model: \Equiv\ differs from \NonEquiv\ in both. \NoSym\ has the circuit of \Equiv, and therefore the same DLA, but a non-equivariant readout. On all four tasks it overfits as much as \NonEquiv\ (Fig.~\ref{fig:main}a--d), and \Equiv\ is more accurate than \NoSym\ in at least $12$ of the $15$ seeds (Fig.~\ref{fig:main}e--h, Table~\ref{tab:paired}). The two non-invariant models have the same test accuracy, except on Galaxy10, where \NoSym\ is less accurate, in part because two of its runs fail to train.

The four models form a $2\times2$ design, the circuit (equivariant or not) crossed with the model (invariant or not), and the effect of each ingredient can be estimated per seed. Making the model invariant reduces the gap and raises the test accuracy on every task, by $0.06$--$0.11$ (Table~\ref{tab:factorial}). Making the circuit equivariant has no consistent effect, as the pairs that differ only in the circuit (\Equiv\ and \Twirled, \NoSym\ and \NonEquiv) show in Table~\ref{tab:paired}: it changes the test accuracy by a few hundredths, in either direction depending on the task, and changes the gap significantly only on the synthetic task, where \Equiv\ overfits less than \Twirled\ ($p=0.03$). Where it helps, on the synthetic task, it does so through the invariant model it makes possible (\Equiv\ beats \Twirled). The equivariance of the circuit thus matters only through the invariant model. Conversely, averaging \NoSym\ over the group turns it back into \Equiv\ (Sec.~\ref{subsec:models}): the equivariant circuit recovers the advantage as soon as the model is invariant.

\subsection{Shortcuts}\label{subsec:results_cues}

\begin{figure*}[tbp]
\centering
\includegraphics[width=\textwidth]{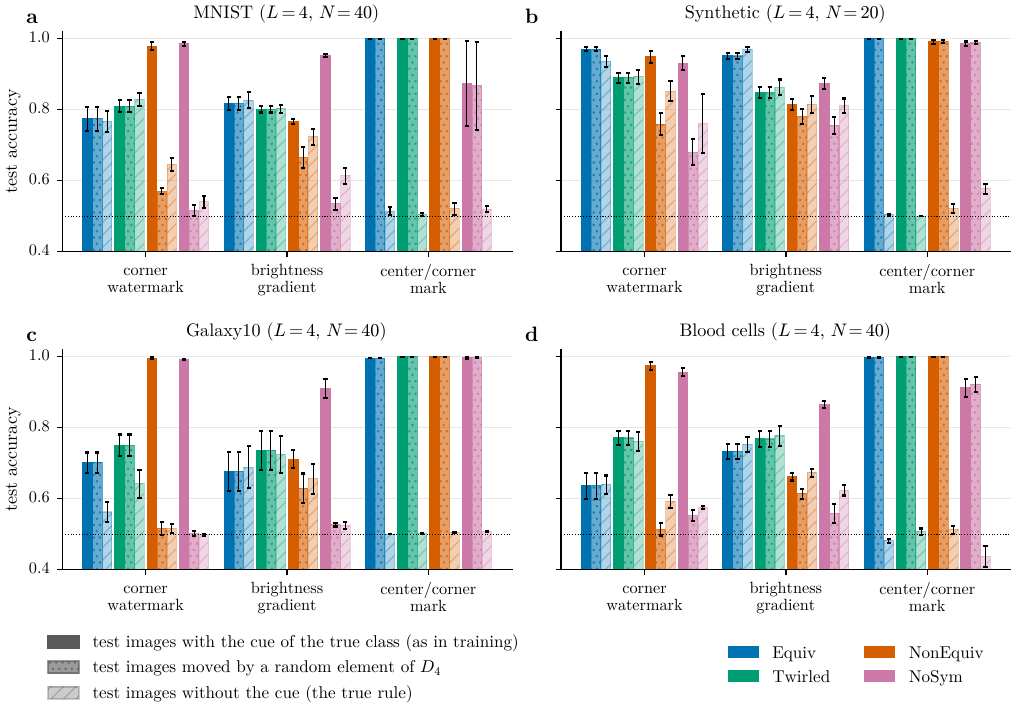}
\caption{Shortcuts: test accuracy of the four models trained with a cue in every training image, in the main settings (five seeds; mean and standard error). Each group of bars is one model, tested on the images with the cue of the true class (solid), on the cued images moved by a random element of $D_4$ (dotted) and on the images without the cue (hatched). The non-invariant models learn the watermark and fail when it is moved; the invariant ones cannot see its position and learn the content of the images. The center/corner mark is invariant under the group, and all the models learn it.}\label{fig:cues}
\end{figure*}

Figure~\ref{fig:cues} shows what the models learn when every training image carries a cue that agrees with its label. The corner watermark, which the group moves, separates the models on every task. The non-invariant models learn it: they are almost perfect on test images that carry the watermark of their class, and drop to chance level when the watermark is moved by the group. On the three tasks with real images they are also close to chance level without the watermark: they have learned the watermark rather than the digit, the galaxy or the cell. On the synthetic task the shapes are easier to learn and the shortcut is only partial. The invariant models cannot tell the two watermarks apart, so moving the watermark leaves their accuracy unchanged, and they learn the content of the images: without the watermark they remain well above the non-invariant models on MNIST, on the synthetic task and on the blood cells. On Galaxy10, however, removing the watermark lowers the accuracy also of the invariant models. They cannot see its position, but they are affected by its weight in the encoded state, a fifth of the squared norm of every image: removing it shifts their output by almost the same amount for both classes, and on Galaxy10, where the outputs of the two classes are close, many test images cross the threshold. Moving the threshold by the shift measured on the training images restores the accuracy obtained with the watermark (analysis and values in the repository, \texttt{experiments/watermark\_shift.py}).

The brightness gradient separates the two non-invariant models. \NoSym\ relies on it almost entirely, while \NonEquiv\ uses it only in part. The readout explains the difference: at the start of training, when the circuit is close to the identity, $Z_0$ and $Z_4$ measure the difference between the squared intensities of the two halves of the image, which is exactly what a brightness gradient changes. The invariant models are unaffected by the gradient.

The control cue, invariant under the group, is learned by all four models, which fall to chance level without it. The invariant models therefore do not ignore the watermark because they cannot see small patches: they ignore the cues that the group moves, and only those.

\subsection{Hardware noise model}\label{subsec:results_noise}

\begin{figure*}[tbp]
\centering
\includegraphics[width=\textwidth]{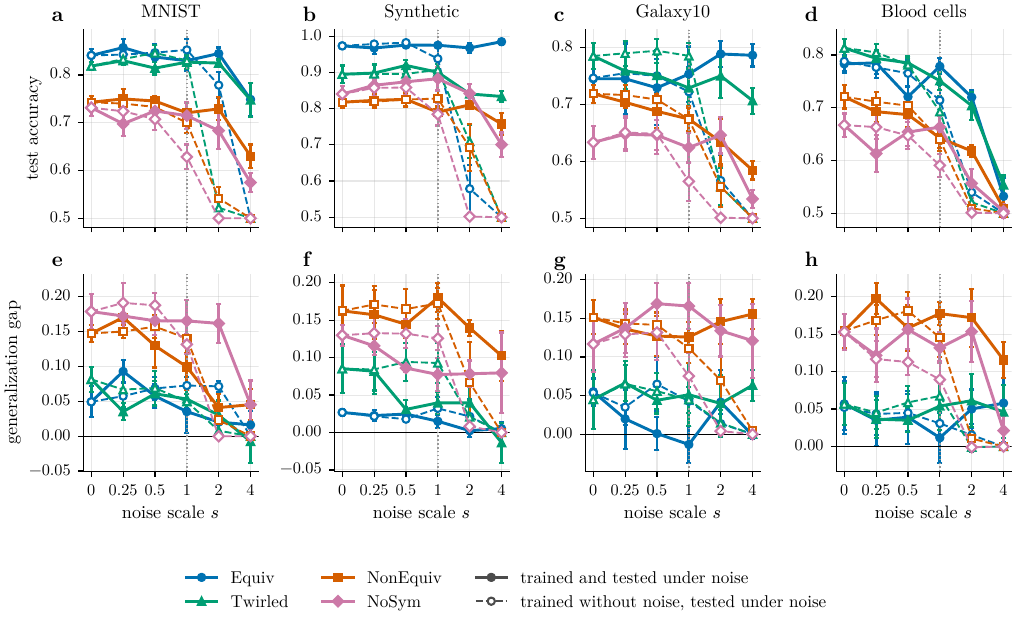}
\caption{Hardware noise: test accuracy (\textbf{a--d}) and generalization gap (\textbf{e--h}) of the four models in the main settings as functions of the noise scale $s$ ($s=0$: noiseless; $s=1$: a typical superconducting processor, dotted line). Solid lines: models trained and tested under noise; dashed lines: models trained without noise and run under noise. Mean and standard error over the five seeds shared by the two scenarios.}\label{fig:noise}
\end{figure*}

Figure~\ref{fig:noise} shows the four models under the noise model of Sec.~\ref{subsec:noise}.

\paragraph{Noise in training and test.} Up to twice the noise of a typical device, the invariant models keep their noiseless accuracy and do not overfit, on every task. The non-invariant models keep overfitting, except \NonEquiv\ on MNIST, whose generalization gap decreases as the noise increases. The decrease comes from the training accuracy, which drops, and not from the test accuracy, which remains approximately constant: the noise prevents \NonEquiv\ from fitting its training images, but does not make it more accurate on unseen ones. The advantage of \Equiv\ over \NonEquiv\ persists at every noise level, and on the synthetic task and on Galaxy10 it grows with the noise. At the strongest noise ($s=4$), the blood cells are the only task on which every model, invariant or not, drops to chance level, so that no advantage is left.

\paragraph{Noise in the test only.} When the models are trained without noise and only tested under noise, the advantage is intact: the noise does not change what the models have learned, and the overfitted ones remain overfitted. \NoSym\ is the most fragile model, because relaxation shifts $\langle Z\rangle$ directly, but not $\langle X\rangle$.

Beyond a typical device every model trained without noise degrades towards a constant prediction: relaxation and the asymmetric readout errors shift the output by a constant that eventually exceeds its variation from image to image. A model trained under noise learns to compensate the shift.

\section{Discussion}\label{sec:discussion}

\paragraph{Why invariance prevents overfitting.} The four models share the gates, the angles and, up to a change of basis, the DLA; none of them has vanishing gradients (Fig.~\ref{fig:gradients}), and they fit their training images comparably (Fig.~\ref{fig:main}a--d). The usual explanations of a difference in generalization, a more expressive circuit \cite{caro2022generalization, abbas2021power} or one that is harder to train \cite{mcclean2018barren, ragone2024lie, larocca2025barren}, therefore do not apply. Nor does the size of the function class: \Twirled\ and \NoSym\ combine a similar number of features, yet the first overfits as little as \Equiv\ and the second as much as \NonEquiv, whose class is almost twice as large.
What decides the overfitting is whether the model is invariant. An invariant model trained on $N$ images is effectively trained on their whole orbits (Sec.~\ref{subsec:terms}), so the pose of an image carries no information for it. A non-invariant model must infer from $N$ images that the pose does not matter, and its features, most of which change with the pose, let it memorize the poses it has seen instead. A partial invariance is not enough: \NonEquiv\ is invariant under the row reflection, so, applied to the subgroup $\{e,t_x\}$, the same argument gives that it effectively sees two of the eight poses of every image, yet it overfits as much as \NoSym, which has no residual symmetry. The upright control confirms the mechanism: once the pose stops varying, the same non-invariant circuit overfits far less. The result does not depend on the data set. It holds on a benchmark, on a synthetic task and on two scientific data sets collected for other purposes, in which the random orientation is a property of the objects themselves.

\paragraph{What the equivariant circuit adds.} An equivariant circuit alone brings no benefit to generalization: with a non-equivariant readout it overfits like a generic circuit. Its role is to give an invariant model at no extra cost, whereas twirling multiplies the number of circuit evaluations by $|G|$, and to fix which invariant model: the function classes of \Equiv\ and \Twirled\ are nearly disjoint and of very different size. Which class is better is a property of the task, as for any choice of model, and both avoid overfitting. One cost of the equivariant circuit is that it sometimes fails to train at large depth, as does \NoSym, which shares it; we have not identified the cause (Appendix~\ref{app:training}).

\paragraph{When the symmetry does not help.} The advantage requires the symmetry to be a property of the data. On upright digits a fixed orientation is informative, and an invariant model discards it: the symmetry restricts what a model can learn, and the restriction helps only when what it excludes carries no information about the label. The advantage also requires a task that the equivariant circuit can represent. In a first version of the synthetic task, with small shapes on a strong background, no circuit learned anything (Appendix~\ref{app:pilots}).

\paragraph{Shortcuts.} Shortcut learning \citep{geirhos2020shortcut,lapuschkin2019unmasking} is an extreme form of the same mechanism: the model relies on a property of the training images that does not determine the label in general. A cue that the transformations of $D_4$ move is invisible to an invariant model, which learns the content of the images regardless of how easy the cue is; a cue that the group does not move is learned by every model. The invariance thus protects against exactly the cues that the group moves. Among the non-invariant models, which cue is preferred depends on the readout.

\paragraph{Hardware noise.} Two facts explain why the advantage survives realistic noise. The invariant models do not rely on fine details of the training images, which the noise blurs; and since all the channels act on single qubits, the noise breaks the symmetry of \Equiv\ only through non-unital relaxation and the differences between qubits \citep{tuysuz2024symmetry}, by $1$--$2\%$ of the output. The readout matters here too: an observable that relaxation shifts directly, such as $Z$, makes a model fragile. Beyond a typical device, the bias of the noisy output makes it necessary to train on the device, or at least to recalibrate the decision threshold, for every model.

\section{Conclusion}\label{sec:conclusion}

We have investigated how incorporating task symmetries affects the
generalization and robustness of quantum neural networks for image
classification. Our study compares four models with matched circuit
layouts, parameter counts and parameter-sharing patterns, using
variational circuits with equivalent dynamical Lie algebras.
This provides a controlled setting for assessing the benefits of
symmetry constraints across four tasks involving handwritten digits,
synthetic shapes, galaxy morphology and red blood cells.

In the random-pose setting, the invariant models achieve smaller
generalization gaps and higher average test accuracy on all four
tasks, despite being trained on only a few tens of images.
The controls on image orientation support the interpretation that
these constraints improve generalization by making the predictions
insensitive to pose-dependent patterns in the training sample.
The shortcut experiments demonstrate a related benefit: the invariant
models resist artificial cues that distinguish the classes through
their orientation or through positions related by the task symmetries.
This protection is specific to the imposed symmetry, since cues that
are themselves invariant can still be learned as shortcuts.
The observed advantage also persists in simulations with noise
parameters representative of superconducting processors, including
thermal relaxation, depolarizing noise and readout errors.

The relative performance of models respecting the same symmetry
nevertheless depends on the task, highlighting the importance of
which invariant functions a chosen architecture can represent.
Architectural constraints and explicit group averaging also differ
in the number of circuit evaluations needed to obtain a prediction.
These results support using known task symmetries to guide quantum
model design when labeled data are limited, while accounting for
representational flexibility and evaluation cost.
They also illustrate the value of controlled benchmarks that match
the parametrization, gate layout and algebraic structure of the
circuits being compared.
Extending this analysis to larger symmetry groups, alternative
encodings and deeper circuits would clarify how these benefits
interact with trainability and computational resources beyond the
regimes studied here.

\section*{Data and code availability}

MNIST \citep{lecun1998gradient}, Galaxy10 SDSS \citep{galaxy10} and erythrocytesIDB \citep{erythrocytesidb} are publicly available, and the synthetic task is generated by the code. The code used to generate the data, train and evaluate the models, compute the dynamical Lie algebras and produce every figure and table, the results of every run, and supplementary tables with every setting not reported here (all numbers of layers and of training images, the shortcuts at every $N$, the noise scenarios at every noise scale and the pilot runs) are available at \url{https://github.com/giuseppedambruoso/eqnn}. The code is released under the GNU General Public License v3.0 or later, and the results, figures and tables under the Creative Commons Attribution-ShareAlike 4.0 International license.

\begin{acknowledgments}

We acknowledge support from INFN through the projects ``QUANTUM'' and ``NPQCD'', and from the Italian funding within the ``Budget MUR - Dipartimenti di Eccellenza 2023-2027'' - Quantum
Sensing and Modelling for One-Health (QuaSiModO). We acknowledge support from Projects PN-RIC Q-SUD (Q-SUD B99H26000380007) and PN-RIC QUANTAS (SYNERGIA B99H26000410007).
We acknowledge computational resources provided by the University of Bari and the INFN cluster ReCaS \cite{RECAS}. GG acknowledges support from the Italian National Group of Mathematical Physics (GNFM-INdAM) and from the project ``Patto territoriale sistema universitario pugliese'' (PATTI).

\textbf{AI tools usage.}
The authors used AI tools (ChatGPT and Claude) during the preparation
of this manuscript to assist with code writing, proofreading and language editing of the manuscript.
All resulting text was reviewed and edited by the authors, who take full responsibility for
the content of the work.

\textbf{SDSS data acknowledgments.}
This work uses images from the Sloan Digital Sky Survey. Funding for the SDSS and SDSS-II has been provided by the Alfred P. Sloan Foundation, the Participating Institutions, the National Science Foundation, the U.S. Department of Energy, the National Aeronautics and Space Administration, the Japanese Monbukagakusho, the Max Planck Society, and the Higher Education Funding Council for England. The SDSS Web Site is \url{http://www.sdss.org/}. The SDSS is managed by the Astrophysical Research Consortium for the Participating Institutions. The Participating Institutions are the American Museum of Natural History, Astrophysical Institute Potsdam, University of Basel, University of Cambridge, Case Western Reserve University, University of Chicago, Drexel University, Fermilab, the Institute for Advanced Study, the Japan Participation Group, Johns Hopkins University, the Joint Institute for Nuclear Astrophysics, the Kavli Institute for Particle Astrophysics and Cosmology, the Korean Scientist Group, the Chinese Academy of Sciences (LAMOST), Los Alamos National Laboratory, the Max-Planck-Institute for Astronomy (MPIA), the Max-Planck-Institute for Astrophysics (MPA), New Mexico State University, Ohio State University, University of Pittsburgh, University of Portsmouth, Princeton University, the United States Naval Observatory, and the University of Washington.
\end{acknowledgments}

\appendix
\begin{table*}[t]
\footnotesize
\begin{tabular*}{\textwidth}{@{\extracolsep{\fill}}llllc@{}}
\toprule
Task & Source & Class 0 & Class 1 & Test images\\
\midrule
MNIST & handwritten digits \citep{lecun1998gradient}, random pose & digit $4$ & digit $5$ & $1000$\\
Synthetic & generated & filled square & thin cross & $1000$\\
Galaxy10 & SDSS images \citep{york2000sloan}, Galaxy Zoo labels \citep{lintott2008galaxy,lintott2011galaxy,willett2013galaxy}, random pose & face-on disk & edge-on disk & $630$\\
Blood cells & erythrocytesIDB, sickle cell disease \citep{gonzalez2015red,erythrocytesidb}, random pose & elongated cell & other deformed cell & $212$\\
\bottomrule
\end{tabular*}
\caption{The four tasks. ``Random pose'': every image, in training and in test, is transformed by an element of $D_4$ drawn uniformly at random, independently of the label. The test sets are balanced.}\label{tab:tasks}
\end{table*}

\section{Dynamical Lie algebra, function classes and gradients}\label{app:dla}\label{app:gradients}

\paragraph{Dynamical Lie algebra.} The DLA is generated by the gates of one layer. The two rotations of a pair share one angle and act simultaneously, so the pair is a single gate with the summed generator. For the equivariant circuit the generators are $X_a+X_{a'}$ ($a=0,\dots,3$, $a'=a+n/2$) and $Y_aY_bY_{a'}Y_{b'}$ for the five blocks $(a,b)\in\{(0,1),(2,3),(1,2),(3,0),(0,2)\}$, nine in all; for the non-equivariant circuit they are $X_a+Y_{a'}$ and $Y_aY_bZ_{a'}Z_{b'}$. Stacking layers adds no generator, so the DLA does not depend on $L$. As is customary, we compute the DLA of the gate set, with an independent angle for every gate. We computed it with the Lie closure of PennyLane (\texttt{qml.lie\_closure}), which takes the nested commutators of the generators, expressed as sums of Pauli strings, until no new linearly independent element appears. For both circuits it gives $\dim\mathfrak g=569$.

\begin{table}[t]
\footnotesize
\begin{tabular*}{\columnwidth}{@{\extracolsep{\fill}}lccccc@{}}
\toprule
Model & DLA & obs. & feat. & inv. & sh.\\
\midrule
\Equiv & 569 & 417 & 237 & 237 & 237 \\
\Twirled & 569 & 1272 & 666 & 666 & 14 \\
\NonEquiv & 569 & 1480 & 1304 & 31 & 4 \\
\NoSym & 569 & 1480 & 772 & 237 & 237 \\
\bottomrule
\end{tabular*}
\caption{Dimensions of the DLA, of the space of observables $\mathcal M_O$ (obs.; for \Twirled, the twirled observables), of the function class (feat.), of its invariant part (inv.) and of the features shared with \Equiv\ (sh.).}\label{tab:dla}
\end{table}

\paragraph{Function classes.} Since the images are real, every model output is $f(x)=x^{\mathsf T}\mathrm{Re}(U^\dagger OU)\,x$, so the size of the function class of a model is the dimension of the span of the matrices $\mathrm{Re}(U^\dagger OU)$ over its circuits. We estimated it as the numerical rank of these matrices for many random circuits, more than the expected dimension, computed in the Heisenberg picture (and group-averaged for \Twirled); the singular values above and below the rank differ by at least thirteen orders of magnitude. With an independent angle per gate the ranks are $237$ for \Equiv, $666$ for \Twirled, $772$ for \NoSym\ and $1304$ for \NonEquiv. With the angles shared as in the trained circuits they are smaller by at most four, less than $0.5\%$ of each class; they are reached with four layers, and two layers already reach almost all of them.

\begin{figure*}[t]
\centering
\includegraphics[width=\textwidth]{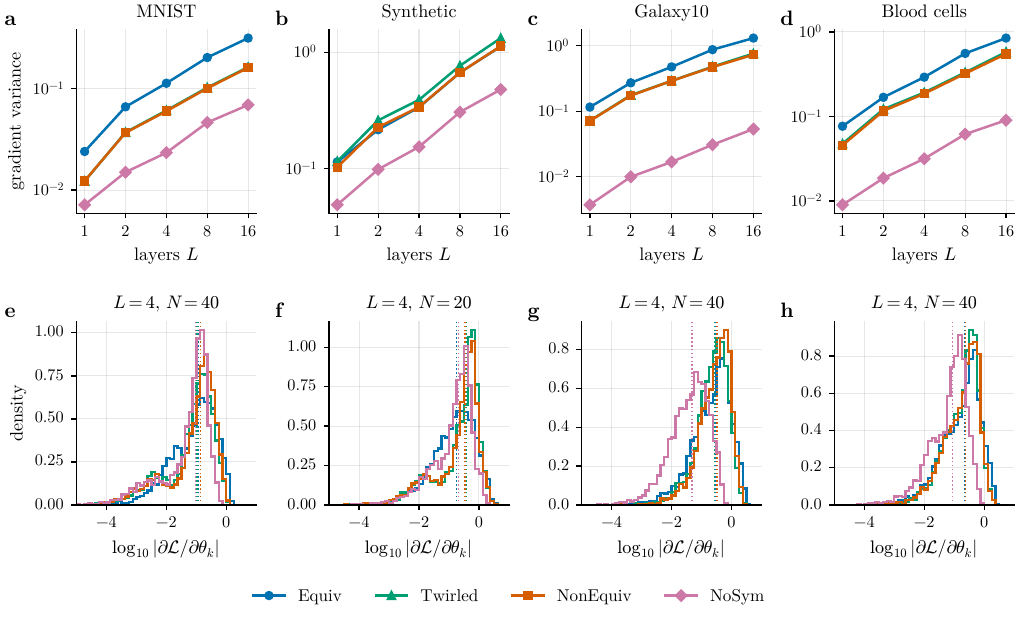}
\caption{ Gradients of the training loss at the initialization used in training (angles uniform in $[-0.1,0.1]$), on the training images of the main settings. \textbf{a--d} Variance over $100$ draws of the angles of each partial derivative, averaged over the angles, as a function of the number of layers. \textbf{e--h} Histograms of the moduli of all the partial derivatives over $200$ draws, with $L=4$; dotted lines: medians.}\label{fig:gradients}
\end{figure*}

\paragraph{Gradients.} To check that none of the models is harder to train than the others, we computed the gradient of the training loss~\eqref{eq:loss} in the regime of our training: angles drawn uniformly in $[-0.1,0.1]$, as at initialization, on the training images of the main settings (\texttt{experiments/gradient\_variance.py} and \texttt{gradient\_histograms.py}). Figure~\ref{fig:gradients} shows the variance of the partial derivatives as a function of the depth, and the distribution of their moduli in the main settings.

No model has vanishing gradients. The variance grows with the depth for all four models, and the moduli of the partial derivatives are spread over two to three orders of magnitude around medians of order $0.1$. \NonEquiv\ and \Twirled\ have almost the same gradients, and \Equiv\ comparable or larger ones; \NoSym\ has the smallest, by up to a factor of six on Galaxy10. Since Adam takes steps that do not depend on the scale of the gradient, these differences do not hinder training, and all four models reach comparable training accuracies. They do not follow the overfitting either: \NoSym\ overfits as much as \NonEquiv\ with smaller gradients, and \Twirled\ as little as \Equiv\ with the gradients of \NonEquiv.

\section{Tasks}\label{app:tasks}

\subsection{Data sets}

Table~\ref{tab:tasks} summarizes the four tasks. All the images are converted to grayscale, brought to $16\times16$ pixels and $\ell_2$-normalized. The color images of Galaxy10 and erythrocytesIDB are first converted to their luminance, $0.299R+0.587G+0.114B$. The images are then downsampled from their original size ($28\times28$ pixels for MNIST, $69\times69$ for Galaxy10 and $80\times80$ for the blood cells) by bilinear interpolation with antialiasing, so that every pixel of the $16\times16$ image is a weighted average of the original pixels around it, and fine details are smoothed rather than aliased. The synthetic images are generated directly at $16\times16$ pixels. 

\paragraph{MNIST.} The training pool is the first $50\,000$ images of the training split, and the test set a fixed balanced set of $1000$ images of the test split. Every image is transformed once by a random element of $D_4$; the control uses the upright images.

\paragraph{Synthetic task.} Each image contains a filled square of side $8$--$14$ pixels (class $0$) or a plus-shaped cross, one pixel thick, with arms of total length $9$--$15$ pixels (class $1$). The size, the position (uniform over the placements inside the frame) and the intensity (uniform in $[0.5,1]$) of the shape are random, and the background pixels are $|\epsilon|$ with $\epsilon\sim\mathcal{N}(0,0.05^2)$. Both shapes are symmetric under $D_4$ about their center and the set of positions is invariant, so the distribution of each class is invariant.

\paragraph{Galaxy10.} Galaxy10 SDSS \citep{galaxy10} contains $21\,785$ color images of galaxies from the Sloan Digital Sky Survey \citep{york2000sloan}, in ten morphological classes from the votes of Galaxy Zoo \citep{lintott2008galaxy,lintott2011galaxy,willett2013galaxy}. We distinguish face-on disk galaxies without spiral arms (class $0$ of the data set) from edge-on disk galaxies with a rounded bulge (class $4$): a round disk from an elongated one, at any orientation. We use the luminance of the images. The data set has no official split, so we split it once at random into $80\%$ for training and $20\%$ for testing; the test set contains all the $315$ edge-on galaxies of the test split and as many face-on ones.

\paragraph{Red blood cells.} erythrocytesIDB \citep{gonzalez2015red,erythrocytesidb} contains $629$ images of single red blood cells from peripheral blood smears of patients with sickle cell disease, classified by a specialist as circular, elongated (the sickle cells) or other (deformed without being sickle-shaped). We distinguish the elongated cells from the other ones. Since the amplitude encoding normalizes the image, the amplitude must lie on the cell rather than on the background: the cells are dark on a bright background, so we use the inverted luminance minus the background level (the median of the border pixels), clipped at zero. The data set is small, and we split every class once at random into two halves, for training and for testing.

\begin{figure*}[t]
\centering
\includegraphics[width=\textwidth]{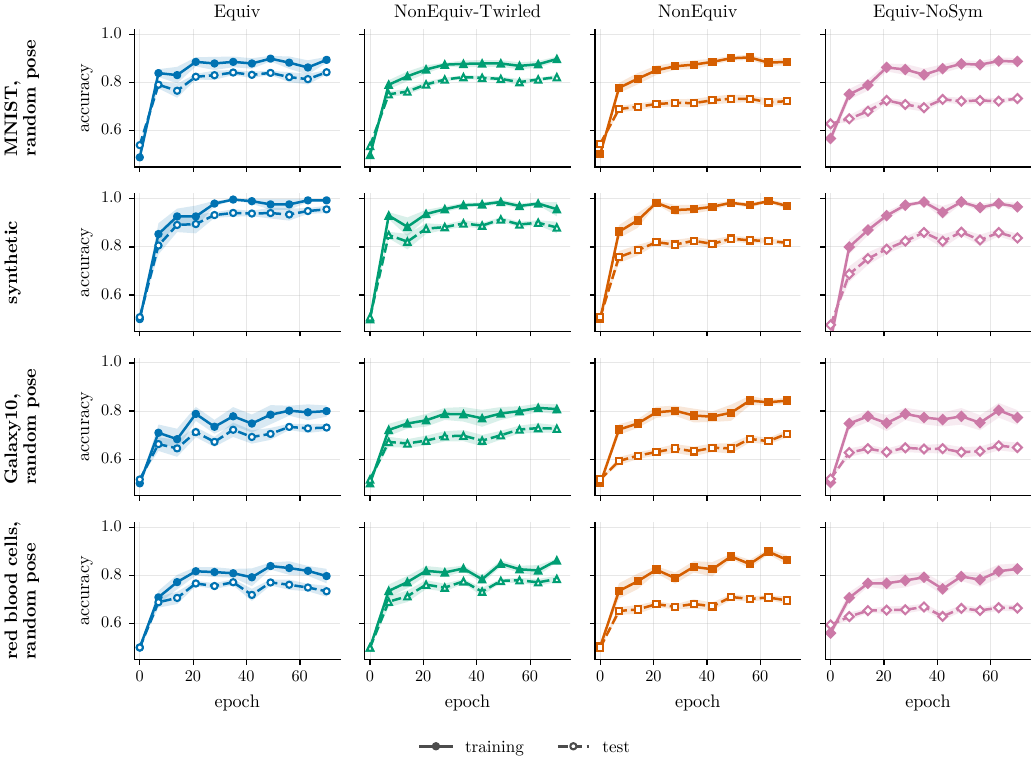}
\caption{Learning curves in the main settings: training accuracy (solid) and accuracy on $250$ test images (dashed) every seven epochs; mean and standard error over $15$ seeds. The test accuracy of the non-invariant models stops growing after a few epochs while their training accuracy keeps rising.}\label{fig:curves}
\end{figure*}

\subsection{Cues}\label{app:cues}

The cues are added to every training image, after the random pose, in the frame of the image. The \emph{corner watermark} is a $2\times2$ patch near a corner, at a reference position for class $0$ and at the position rotated by $90^\circ$ for class $1$; its amplitude is set so that it carries $20\%$ of the squared norm of every image, so that neither its intensity nor its weight in the encoded state depends on the class. The \emph{brightness gradient} multiplies the image by a ramp that changes by a factor of ten along an oblique direction, rotated by $90^\circ$ for class $1$. The \emph{center/corner mark} is a patch of twice the maximal intensity of the image, at the center of the frame for class $0$ and in its four corners for class $1$. Since the content of the images is in a random pose, the orbit of a cued image does not reveal which of the two watermarks or gradients it carries.

\subsection{Choice of the tasks}\label{app:pilots}

The parameters of the synthetic task and the pairs of classes of Galaxy10 and erythrocytesIDB were chosen in short pilot runs with \Equiv\ and \NonEquiv\ only, at two depths and with three seeds. The pilot results are in the repository.

\paragraph{Synthetic task.} The first version had small shapes on a strong background noise, and no circuit learned it. At the start of training the circuits are close to the identity, and the observable $(X_0+X_4)/2$ measures the correlation between each pixel and the pixel half an image away in the same row or column, since qubits $0$ and $4$ hold the most significant bits of the indices. A small shape gives no such correlation, so the training starts without any signal about it. We therefore made the shapes large and the noise weak, and compared a dozen variants that differ in the range of the sizes, the thickness of the cross and the strength of the noise. \Equiv\ overfit less than \NonEquiv\ in all of them, and was more accurate in all but one (hollow squares), where the two circuits were equally accurate. Smaller shapes increased the overfitting of both circuits, and a thicker cross or more noise reduced the difference between them. We chose the variant with the thinnest cross, on which \Equiv\ is most accurate; the variants closest to it give similar results.

\paragraph{Galaxy10.}
\begin{figure*}[t]
\centering
\includegraphics[width=\textwidth]{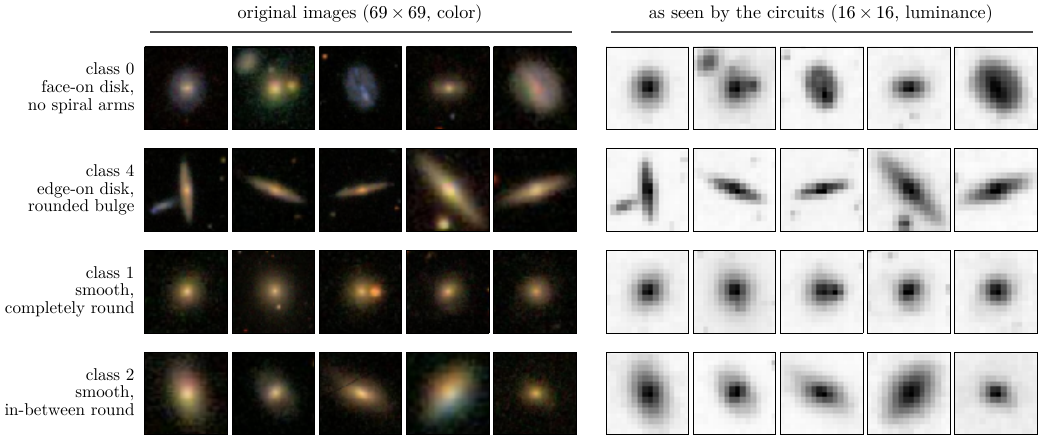}
\caption{The four classes of Galaxy10 considered in the pilot runs; the task of the paper is class $0$ against class $4$. Left: the first five images of each class in the data set ($69\times69$ pixels, color). Right: the same images as seen by the circuits, converted to luminance and downsampled to $16\times16$ pixels.}\label{fig:galaxy_classes}
\end{figure*} 
We considered five pairs of the classes of Galaxy10 \citep{galaxy10} whose distinction rests on the shape of the galaxy and not on its orientation (Fig.~\ref{fig:galaxy_classes}): face-on disk without spiral arms (class $0$) against edge-on disk with a rounded bulge (class $4$); smooth, completely round galaxies (class $1$) against classes $0$ and $4$; smooth, in-between round galaxies (class $2$) against class $4$; and class $1$ against class $2$.

\paragraph{Red blood cells.} Of the three possible pairs, circular against elongated cells is so easy that both circuits learn it without overfitting, so there is little overfitting for the invariance to prevent; circular against other cells is separated poorly by both circuits. We chose the remaining pair, elongated against other cells.

\section{Training and evaluation}\label{app:training}\label{app:scaling}

\paragraph{Simulation and loss.} The circuits are simulated exactly, in double precision, with PennyLane (\texttt{default.qubit}) and backpropagation in PyTorch \citep{bergholm2018pennylane,paszke2019pytorch}. The state $\ket{\psi(x)}$ is prepared exactly. We minimize the logistic loss
\begin{equation}
\mathcal{L}=\frac1N\sum_{i=1}^N\log\bigl(1+e^{-\beta \tilde y_i f(x_i)}\bigr),\qquad \tilde y_i=2y_i-1,
\label{eq:loss}
\end{equation}
with $\beta=20$.
Since $f$ is bounded in $[-1,1]$, with $\beta=1$ the margin $\tilde y_if(x_i)$, positive for a correctly classified image, would stay in the region where $\log(1+e^{-z})\approx\log 2-z/2$ is almost linear, and every image would contribute almost equally to the gradient, whatever its margin. With $\beta=20$ the loss saturates for images classified correctly with a margin larger than a few times $1/\beta$, and grows linearly for misclassified ones, as a smooth hinge loss; a fixed scale of this kind is commonly applied to bounded outputs in classical classifiers \citep{wang2017normface}. The scale enters the training only, not the prediction.

\paragraph{Optimization.} The angles are initialized uniformly in $[-0.1,0.1]$ and trained with Adam \citep{kingma2015adam}, 
learning rate $0.05$ and decay rates $0.5$ and $0.999$ for the first and second moment estimates, for $70$ epochs of ten mini-batches of $N/10$ images (four images for $N=40$, two for $N=20$). A seed fixes the training images, the random poses, the initial angles and the order of the mini-batches. The learning curves (Fig.~\ref{fig:curves}) evaluate the test accuracy on $250$ test images every seven epochs.

\paragraph{Grid and statistics.} We ran $L\in\{1,2,4,8\}$ layers ($L\in\{1,2,4\}$ on the synthetic task) and $N\in\{10,20,40,80\}$ training images with five seeds each, and added ten seeds in the main settings, with $L\in\{2,4\}$. The paper reports $L=4$; the results with $L=2$ are similar and are in the repository. Error bars are standard errors over the seeds. Comparisons use the seeds shared by all the compared models and the two-sided Wilcoxon signed-rank test; we also report the number of seeds in which a difference is positive. The experiments on shortcuts use five seeds per setting.

\begin{table*}[t]
\footnotesize
\begin{tabular*}{\textwidth}{@{\extracolsep{\fill}}ll ccc ccc@{}}
\toprule
 & & \multicolumn{3}{c}{Test accuracy} & \multicolumn{3}{c}{Reduction of the gap}\\
\cmidrule(lr){3-5}\cmidrule(l){6-8}
Task & Pair & mean & $k$ & $p$ & mean & $k$ & $p$\\
\midrule
MNIST & Equiv vs NonEquiv & $+0.11(1)$ & 15 & $7\times10^{-4}$ & $+0.10(2)$ & 14 & $2\times10^{-3}$ \\
 & Twirled vs NonEquiv & $+0.10(1)$ & 15 & $7\times10^{-4}$ & $+0.09(2)$ & 13 & $3\times10^{-4}$ \\
 & Equiv vs Twirled & $+0.01(1)$ & 9 & $0.24$ & $+0.01(1)$ & 8 & $0.60$ \\
 & Equiv vs NoSym & $+0.12(1)$ & 15 & $7\times10^{-4}$ & $+0.11(2)$ & 14 & $1\times10^{-3}$ \\
 & NoSym vs NonEquiv & $-0.01(1)$ & 6 & $0.44$ & $-0.01(2)$ & 6 & $0.68$ \\
\midrule
Synthetic & Equiv vs NonEquiv & $+0.14(1)$ & 14 & $8\times10^{-4}$ & $+0.12(2)$ & 15 & $7\times10^{-4}$ \\
 & Twirled vs NonEquiv & $+0.07(2)$ & 14 & $8\times10^{-3}$ & $+0.08(2)$ & 13 & $2\times10^{-3}$ \\
 & Equiv vs Twirled & $+0.07(3)$ & 13 & $7\times10^{-3}$ & $+0.04(1)$ & 12 & $0.03$ \\
 & Equiv vs NoSym & $+0.13(2)$ & 14 & $8\times10^{-4}$ & $+0.11(1)$ & 15 & $7\times10^{-4}$ \\
 & NoSym vs NonEquiv & $+0.01(2)$ & 10 & $0.52$ & $+0.01(2)$ & 9 & $0.72$ \\
\midrule
Galaxy10 & Equiv vs NonEquiv & $+0.03(1)$ & 12 & $0.02$ & $+0.07(2)$ & 13 & $6\times10^{-4}$ \\
 & Twirled vs NonEquiv & $+0.03(1)$ & 10 & $0.09$ & $+0.06(2)$ & 11 & $3\times10^{-3}$ \\
 & Equiv vs Twirled & $+0.00(2)$ & 8 & $0.93$ & $+0.01(2)$ & 9 & $0.76$ \\
 & Equiv vs NoSym & $+0.09(1)$ & 15 & $7\times10^{-4}$ & $+0.07(2)$ & 12 & $2\times10^{-3}$ \\
 & NoSym vs NonEquiv & $-0.07(1)$ & 0 & $6\times10^{-4}$ & $+0.00(2)$ & 8 & $0.82$ \\
\midrule
Blood cells & Equiv vs NonEquiv & $+0.04(3)$ & 12 & $0.12$ & $+0.11(2)$ & 14 & $6\times10^{-4}$ \\
 & Twirled vs NonEquiv & $+0.09(1)$ & 14 & $8\times10^{-4}$ & $+0.09(2)$ & 14 & $1\times10^{-3}$ \\
 & Equiv vs Twirled & $-0.05(2)$ & 2 & $0.03$ & $+0.01(2)$ & 9 & $0.49$ \\
 & Equiv vs NoSym & $+0.07(3)$ & 12 & $0.03$ & $+0.10(2)$ & 13 & $9\times10^{-4}$ \\
 & NoSym vs NonEquiv & $-0.03(2)$ & 6 & $0.16$ & $+0.00(2)$ & 8 & $0.80$ \\
\bottomrule
\end{tabular*}
\caption{Paired comparisons in the main settings ($L=4$, $15$ seeds). For the pair ``A vs B'': difference in test accuracy (A minus B) and reduction of the gap (gap of B minus gap of A), mean with the standard error on the last digit; $k$: seeds in which the difference is positive; $p$: two-sided Wilcoxon signed-rank test.}\label{tab:paired}
\end{table*}

\paragraph{Choice of the main settings.}

\begin{figure*}[t]
\centering
\includegraphics[width=\textwidth]{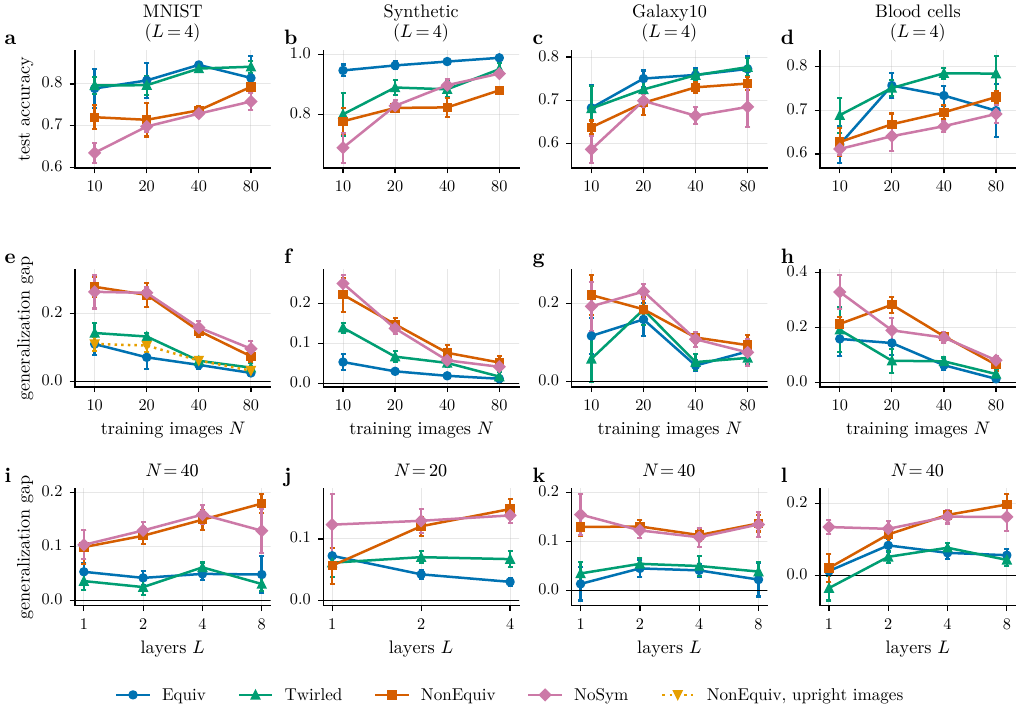}
\caption{Dependence on the number of training images and on the depth. \textbf{a--d} Test accuracy and \textbf{e--h} generalization gap as functions of the number of training images, with $L=4$; the yellow dotted line in \textbf{e} is \NonEquiv\ trained and tested on upright digits. \textbf{i--l} Generalization gap as a function of the number of layers, with the $N$ of the main setting. Mean and standard error over five seeds ($15$ in the main settings).}\label{fig:scaling}
\end{figure*}

The main settings were chosen after running the whole grid, by three criteria. First, enough layers for the non-equivariant circuit to fit its training images: with one layer it fits them poorly, and the comparison would partly measure underfitting rather than overfitting (Fig.~\ref{fig:scaling}i--l). Second, a depth at which the equivariant circuit trains reliably, which excludes eight layers (see below). Third, a number of images at which the training accuracies of the four models are close, so that the gap compares generalization rather than fitting: $N=40$ on MNIST, and $N=20$ on the synthetic task, which is easier to fit. On Galaxy10 and on the blood cells, whose pairs of classes were chosen at $N=40$, we kept that value.

\paragraph{Failures to train.} We call a failure a run whose training accuracy stays at or below $0.6$. Failures occur mostly with eight layers, where \Equiv\ fails in about a third of the runs on MNIST, Galaxy10 and the blood cells, and \NoSym, which shares its circuit, in half of the runs on Galaxy10 and the blood cells; \NonEquiv\ and \Twirled\ rarely fail, except \Twirled\ with eight layers on Galaxy10 (a quarter of the runs). In the main settings there are four failures in total (one of \Equiv\ on the blood cells, two of \NoSym\ on Galaxy10 and one of \Twirled\ on the synthetic task). The counts are in the repository.

\paragraph{Number of images and depth.} Figure~\ref{fig:scaling} shows how the results depend on the number of training images and on the depth. The gap of the non-invariant models is largest with few images and shrinks as $N$ grows, but stays above that of the invariant models, and their test accuracy approaches that of the invariant models only slowly: on MNIST, \NonEquiv\ trained on $80$ images is still less accurate than \Equiv\ trained on $40$. On the synthetic task \NoSym, whose function class contains that of \Equiv, eventually finds functions close to the invariant ones as $N$ grows. With one layer the non-equivariant circuit fits its training images poorly; on MNIST, on the synthetic task and on the blood cells its gap is then smaller, and grows with $L$, as the depth gives it the capacity to memorize. On Galaxy10 it overfits already with one layer. The gap of the invariant models does not grow with the depth. Comparison~1 over the whole grid, with the values of every setting, is in the repository.

\paragraph{Learning curves.} Figure~\ref{fig:curves} shows how the gap forms during training in the main settings. The test accuracy of the non-invariant models stops growing after a few epochs while their training accuracy keeps rising, whereas for the invariant models the two grow together.

\paragraph{Values of the main settings.} Table~\ref{tab:paired} gives the paired comparisons of Fig.~\ref{fig:main}e--h; the training accuracy, test accuracy and generalization gap of Fig.~\ref{fig:main}a--d are in the repository.

\section{Noise model}\label{app:noise}

\begin{figure*}[t]
\centering
\includegraphics[width=\textwidth]{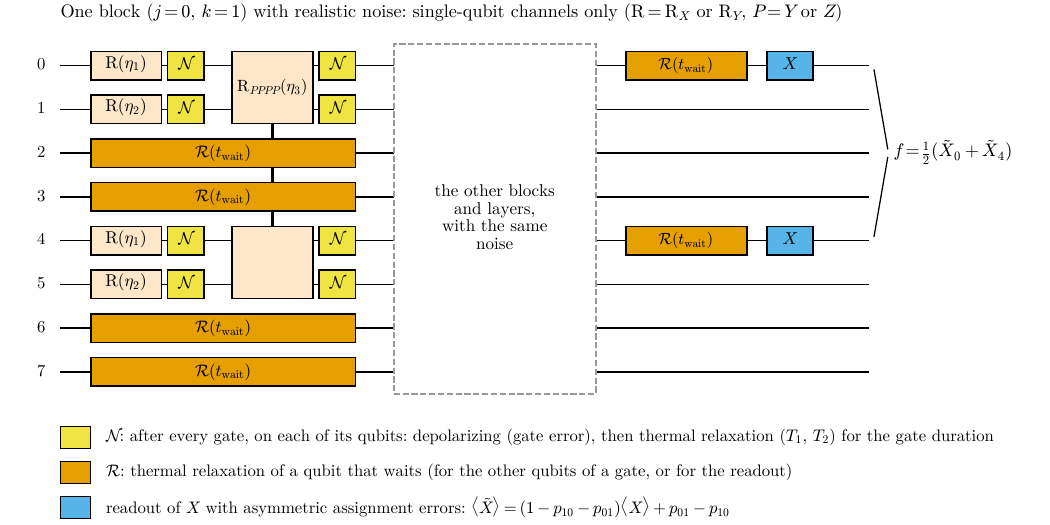}
\caption{The noise model on one block of the circuit. After every gate, each of its qubits undergoes depolarizing noise and thermal relaxation for the duration of the gate ($\mathcal N$); qubits that wait relax for the waiting time ($\mathcal R$); the measured qubits have asymmetric readout errors. All the channels act on single qubits, with parameters that differ from qubit to qubit. The four-qubit rotation $\mathrm R_{PPPP}(\eta_3)$ acts on qubits $0$, $1$, $4$ and $5$; it is drawn as two boxes joined by a thick vertical line, which crosses qubits $2$ and $3$ without acting on them.}\label{fig:noise_model}
\end{figure*}

\paragraph{Channels.} The noise model consists of single-qubit channels with the typical parameters of superconducting processors (Table~\ref{tab:noise_params}, Fig.~\ref{fig:noise_model}). The gates are those of Fig.~\ref{fig:circuits}, not compiled into native gates, and scheduled as soon as possible: a gate starts when all its qubits are free, and the qubits exchanged by the swap start and end their gates at the same times.
After every gate, each of its qubits undergoes the depolarizing channel
\begin{equation}
\mathcal D_\lambda(\rho)=(1-\lambda)\,\rho+\lambda\,\tfrac{\mathbb I}{2}\tr\rho
\end{equation}
and then thermal relaxation for the duration $t$ of the gate \citep{nielsen2010quantum}: amplitude damping with $\gamma=1-e^{-t/T_1}$, followed by pure dephasing with $\gamma_\phi=1-e^{-2t/T_\phi}$, where $1/T_\phi=1/T_2-1/(2T_1)$, so that the coherences decay as $e^{-t/T_2}$ (the Kraus operators are given below). As in the device noise models of Qiskit Aer \citep{javadiabhari2024qiskit}, the strength $\lambda$ is chosen so that the average infidelity of the noise on the qubit equals the error $\varepsilon$ of the gate. A single-qubit channel with entanglement fidelity $F_e$ has average infidelity $1-(2F_e+1)/3$ \citep{nielsen2002simple}, which is $\lambda/2$ for $\mathcal D_\lambda$; since to first order the infidelities of the two channels add, $\lambda=2(\varepsilon-r_{\mathrm{rel}})$, with $r_{\mathrm{rel}}$ the average infidelity of the relaxation (and $\lambda=0$ if $r_{\mathrm{rel}}\ge\varepsilon$). A qubit that waits, for the other qubits of a gate or for the readout, relaxes for the waiting time. The measured qubits have asymmetric readout errors, which turn a measured $\langle P\rangle$ into $(1-p_{10}-p_{01})\langle P\rangle+p_{01}-p_{10}$. The parameters of each qubit are drawn once from log-normal distributions around the typical values, and the global scale $s$ multiplies every error rate and relaxation rate. We use log-normal distributions because these parameters are positive and vary from qubit to qubit by multiplicative factors rather than by fixed amounts; the median and the standard deviation of the logarithm (Table~\ref{tab:noise_params}) fix the typical value and the relative spread. The state preparation is ideal, and for \Twirled\ the transformations of the input remain classical and noiseless.

\paragraph{Noise of the four-qubit rotations.} A rotation $\mathrm R_P(\eta)=e^{-i\eta P/2}$ on four qubits, with $P$ a Pauli string such as $Y_jY_kY_{j+n/2}Y_{k+n/2}$, is applied as an ideal unitary, followed on each of its four qubits by the same single-qubit channels as a single-qubit gate, with the parameters of its usual compilation into two-qubit gates: an error of $9\times10^{-3}$ per qubit (three two-qubit gates of error $3\times10^{-3}$) and a duration of $480$\,ns (six layers of two-qubit gates). The model therefore contains no correlated multi-qubit error: the noise of a four-qubit gate is the product of four independent single-qubit channels, each with the parameters of its own qubit.

\paragraph{Heisenberg picture.} The noisy circuit is a sequence, in time order, of gates and single-qubit channels, $\Phi=\mathcal E_T\circ\dots\circ\mathcal E_1$, and the output of a model is
\begin{equation}
\begin{split}
f(x)&=\tr\bigl[O\,\Phi(\rho(x))\bigr]=\tr\bigl[\Phi^\dagger(O)\,\rho(x)\bigr],\\
\Phi^\dagger&=\mathcal E_1^\dagger\circ\dots\circ\mathcal E_T^\dagger ,
\end{split}
\label{eq:heisenberg}
\end{equation}
where $\mathcal E^\dagger$, defined by $\tr[A\,\mathcal E(\rho)]=\tr[\mathcal E^\dagger(A)\,\rho]$, is the adjoint of $\mathcal E$. We start from the measured observable, in which the readout errors are already included (each measured Pauli operator $P$ is replaced by $(1-p_{10}-p_{01})P+(p_{01}-p_{10})\mathbb I$), and apply the adjoints of the operations in reverse time order. The adjoints are:
\begin{itemize}
\item for a gate $U$, $\mathcal E^\dagger(A)=U^\dagger AU$;
\item for a channel with Kraus operators $\{K_j\}$, $\mathcal E^\dagger(A)=\sum_jK_j^\dagger AK_j$; amplitude damping has $K_0=\mathrm{diag}(1,\sqrt{1-\gamma})$ and $K_1=\sqrt\gamma\,\ket0\bra1$, pure dephasing $K_0=\mathrm{diag}(1,\sqrt{1-\gamma_\phi})$ and $K_1=\sqrt{\gamma_\phi}\,\ket1\bra1$;
\item the depolarizing channel on qubit $w$ is self-adjoint, $\mathcal E^\dagger(A)=(1-\lambda)A+\lambda\,\frac{\mathbb I_w}{2}\otimes\tr_w A$.
\end{itemize}

\begin{table*}[t]
\begin{minipage}[t]{0.48\textwidth}
\caption{Parameters of the noise model at $s=1$: medians of a typical superconducting processor and log-normal dispersion from qubit to qubit (standard deviation of the logarithm). A single-qubit rotation is two $\sqrt{X}$ pulses; a rotation on four qubits is charged, on each qubit, three two-qubit gates of error $3\times10^{-3}$ and the six layers ($80$\,ns each) of its compilation.}\label{tab:noise_params}
\footnotesize
\begin{tabular*}{\linewidth}{@{\extracolsep{\fill}}lcc@{}}
\toprule
Parameter & Median & Dispersion\\
\midrule
Error, 1-qubit rotation & $4\times10^{-4}$ & $0.5$\\
Duration, 1-qubit rotation & $100$\,ns & --\\
Error, 4-qubit rotation (per qubit) & $9\times10^{-3}$ & $0.4$\\
Duration, 4-qubit rotation & $480$\,ns & --\\
$T_1$ & $200\,\mu$s & $0.3$\\
$T_2$ ($\le 2T_1$) & $150\,\mu$s & $0.3$\\
Readout $p_{10}=P(1|0)$ & $0.01$ & $0.5$\\
Readout $p_{01}=P(0|1)$ & $0.02$ & $0.5$\\
\bottomrule
\end{tabular*}
\end{minipage}\hfill
\begin{minipage}[t]{0.48\textwidth}
\caption{Breaking of the symmetry of \Equiv\ by the noise: mean of $|f(x)-f(R_\mathcal{X}(g)x)|$ over $200$ MNIST test images in random pose, relative to the mean of $|f(x)|$, for angles drawn uniformly in $[-0.5,0.5]$ (ten draws), $L=4$, with identical qubits or with the dispersion of Table~\ref{tab:noise_params}. With identical qubits the transposition remains an exact symmetry.}\label{tab:symbreak}
\footnotesize
\setlength{\tabcolsep}{2.5pt}
\begin{tabular*}{\linewidth}{@{\extracolsep{\fill}}lc cccc@{}}
\toprule
Qubits & $s$ & transp. & rot.\ $90^\circ$ & refl. & rot.\ $180^\circ$\\
\midrule
identical & 0.25 & 0.0\% & 0.2\% & 0.2\% & 0.3\% \\
 & 1 & 0.0\% & 0.6\% & 0.6\% & 0.8\% \\
 & 4 & 0.0\% & 0.4\% & 0.4\% & 0.6\% \\
dispersion & 0.25 & 0.4\% & 0.5\% & 0.2\% & 0.3\% \\
 & 1 & 1.5\% & 1.7\% & 0.6\% & 0.8\% \\
 & 4 & 2.3\% & 2.4\% & 0.3\% & 0.4\% \\
\bottomrule
\end{tabular*}
\end{minipage}
\end{table*}

\paragraph{Symmetry breaking.} A depolarizing channel of the same strength on every qubit would keep \Equiv\ exactly invariant: it commutes with every single-qubit unitary, hence with the reflections, and the swap maps the noise on a qubit to the same noise on its partner. Realistic noise breaks the symmetry in two ways. Amplitude damping is not unital and does not commute with $X$, which breaks the reflections \citep{tuysuz2024symmetry}; and the qubits exchanged by the swap have different parameters, which breaks the transposition. Both effects are small (Table~\ref{tab:symbreak}).


\bibliography{references}

\end{document}